\documentclass[12pt]{article}
\usepackage{bbm}
\usepackage{color}
\usepackage{authblk}
\usepackage{latexsym}
\usepackage{epsfig,amssymb,euscript, mathrsfs,cite}
\usepackage{amsmath}
\usepackage{mathrsfs} 
\usepackage{enumerate}
\definecolor{MyDarkBlue}{rgb}{0.15,0.15,0.45}
\usepackage[linktocpage=true]{hyperref}
\hypersetup{
colorlinks=true,
citecolor=MyDarkBlue,
linkcolor=MyDarkBlue,
urlcolor=MyDarkBlue,
pdfauthor={},
pdftitle={},
pdfsubject={hep-th}
}
\def\v{\vec{v}}
\def\w{\vec{w}}
\def\u{\vec{u}}

\newsavebox{\ns}
\newsavebox{\dbrane}
\newsavebox{\dbshort}

\def\be{\begin{equation}}
\def\ee{\end{equation}}
\def\bea{\begin{eqnarray}}
\def\eea{\end{eqnarray}}

\newcommand{\nn}{\nonumber}

\newcommand\R{\mathbb{R}}
\newcommand\Z{\mathbb{Z}}

\newcommand\C{\mathbb{C}}

\newcommand\diff{\mathrm{d}}

\newcommand{\dd}{\mathrm{d}}
\newcommand{\me}{\mathrm{e}}
\newcommand{\ii}{\mathrm{i}}

\newcommand{\ex}{\mathrm{e}}
\newcommand{\vol}{\mathrm{vol}}

\newcommand{\Ypq}{\mathscr{Y}^{p,q}}
\newcommand{\p}{p}
\newcommand{\q}{q}
\newcommand{\Labc}{\mathscr{L}^{\mathtt{a},\mathtt{b},\mathtt{c}}}
\newcommand{\zz}{\gamma}
\newcommand{\betabeta}{B}

\newcommand{\aaaa}{\mathtt{a}}
\newcommand{\bbbb}{\mathtt{b}}
\newcommand{\cccc}{\mathtt{c}}
\newcommand{\dddd}{\mathtt{d}}

\newcommand{\ppp}{{p}}
\newcommand{\qqq}{{q}}

\newcommand{\sF}{\mathscr{F}}

\newcommand{\cZ}{\mathscr{Z}}
\newcommand{\csugra}{c_{\mathrm{sugra}}}

\newcommand{\etpsi}{{\varphi}}

\newcommand{\BB}{B}
\newcommand{\cc}{c}
\newcommand{\Ssusy}{S_{\mathrm{SUSY}}}
\newcommand{\z}{z}
\newcommand{\mm}{\mathtt{m}}

\newlength{\sswidth}

\numberwithin{equation}{section}       

\begin{document}

\begin{titlepage}

\begin{flushright}
Imperial/TP/2018/JG/03\\
\end{flushright}

\vskip 1cm

\begin{center}


{\Large \bf A geometric dual of $c$-extremization}

\vskip 1cm

{Christopher Couzens$^{\mathrm{a}}$, Jerome P. Gauntlett$^{\mathrm{b}}$, 
Dario Martelli$^{\mathrm{c},\dagger}$\renewcommand*{\thefootnote}{\fnsymbol{footnote}}
\footnotetext[2]{On leave at the Galileo Galilei Institute, Largo Enrico Fermi, 2, 50125 Firenze, Italy.}and James Sparks$^{\mathrm{d}}$}

\vskip 0.5cm
${}^{\mathrm{a}}$\textit{Institute for Theoretical Physics and Center for Extreme\\ Matter
and Emergent Phenomena, Utrecht University,\\
Princetonplein 5, 3584 CE Utrecht, The Netherlands\\}

\vskip 0.2cm

${}^{\mathrm{b}}$\textit{Blackett Laboratory, Imperial College, \\
Prince Consort Rd., London, SW7 2AZ, U.K.\\}

\vskip 0.2cm
${}^{\mathrm{c}}$\textit{Department of Mathematics, King's College London, \\
The Strand, London, WC2R 2LS,  U.K.\\}

\vskip 0.2 cm
${}^{\,\mathrm{d}}$\textit{Mathematical Institute, University of Oxford,\\
Andrew Wiles Building, Radcliffe Observatory Quarter,\\
Woodstock Road, Oxford, OX2 6GG, U.K.\\}

\end{center}

\vskip 0.5 cm

\begin{abstract}
\noindent  We consider supersymmetric AdS$_3 \times Y_7$ and AdS$_2 \times Y_9$ 
solutions of type IIB and $D=11$ supergravity, respectively, that are holographically dual to SCFTs
with $(0,2)$ supersymmetry in two dimensions and $\mathcal{N}=2$ supersymmetry in one dimension.
The geometry of $Y_{2n+1}$, which can be defined for $n\ge 3$,
shares many similarities with Sasaki-Einstein geometry, including the existence of a canonical
R-symmetry Killing vector, but there are also some crucial differences. 
We show that the R-symmetry Killing vector may be determined by 
extremizing a function that depends only on certain global, topological data. In particular, 
assuming it exists, for $n=3$ one can compute the central charge of an AdS$_3 \times Y_7$ 
solution without knowing its explicit form. We interpret this as a geometric dual of $c$-extremization in $(0,2)$ SCFTs.
For the case of AdS$_2 \times Y_9$ solutions we show that the extremal problem can be used to obtain
properties of the dual quantum mechanics, including obtaining the entropy of a class of supersymmetric black holes in AdS$_4$.
We also study many specific examples of the type AdS$_3\times T^2 \times Y_5$, including a new family of explicit supergravity solutions.  In addition we discuss the possibility that the $(0,2)$ SCFTs dual to these solutions can arise from
the compactification on $T^2$ of certain $d=4$ quiver gauge theories associated with five-dimensional Sasaki-Einstein metrics and, surprisingly, come to a negative conclusion.

\end{abstract}

\end{titlepage}

\pagestyle{plain}
\setcounter{page}{1}
\newcounter{bean}
\baselineskip18pt
\tableofcontents

\newpage

\section{Introduction}\label{sec:intro}

The class of supersymmetric conformal field theories (SCFTs) in two spacetime dimensions that preserve $(0,2)$ supersymmetry
share several similarities with the class of ${\cal N}=1$ SCFTs in four spacetime dimensions.
For example, in both cases the SCFTs possess a continuous abelian R-symmetry which determines
exact results about the spectrum of operators. Furthermore, the R-symmetry can be obtained,
in rather general circumstances, by solving a variational problem whereby one extremizes a certain functional over the space of possible R-symmetries. In $d=2$ this procedure, known as $c$-extremization \cite{Benini:2012cz}, also yields the right moving central charge, $c_R$, of the SCFT. This is the direct analogue of the procedure in $d=4$, known as $a$-maximization \cite{Intriligator:2003jj}, which yields the $a$ central charge. 

A principal aim of this paper is to present a geometric dual\footnote{We note that a connection between $c$-extremization and three-dimensional gauged supergravity was made in
\cite{Karndumri:2013iqa}, generalizing a similar connection between $a$-maximization  and five-dimensional gauged supergravity in \cite{Tachikawa:2005tq}.} of $c$-extremization for the class of $d=2$, $(0,2)$ SCFTs with holographic duals given by AdS$_3\times Y_7$ solutions of type IIB supergravity with only five-form flux \cite{Kim:2005ez}. This result can be viewed as the analogue of the geometric dual of $a$-maximization that was presented in 
\cite{Martelli:2005tp,Martelli:2006yb}, for the general class of $d=4$, ${\cal N}=1$ SCFTs which are holographically
dual to AdS$_5\times SE_5$ solutions of type IIB supergravity, where $SE_5$ is a five-dimensional Sasaki-Einstein manifold.

Before discussing the AdS$_3\times Y_7$ solutions of interest further, we first recall some salient aspects of the story involving AdS$_5\times SE_5$ solutions. These solutions, which also have only  
five-form flux, arise from D3-branes sitting at the apex of the Calabi-Yau cone whose link ({\it i.e.} cross section) is the $SE_5$ manifold. The R-symmetry of the dual field theory is geometrically realized as a nowhere vanishing ``Reeb'' Killing vector on $SE_5$. In the geometric dual of $a$-maximization discussed in \cite{Martelli:2005tp,Martelli:2006yb}, one first goes ``off-shell'' by fixing the complex structure of the cone and then considering the more general class of compatible Sasaki metrics on the link of this cone. It is only when the Sasaki metric is also Einstein that the type IIB equations of motion are satisfied. 
It was shown that the Reeb vector field for the Sasaki-Einstein metric can be obtained by minimizing the
normalized volume of the Sasaki manifold as a functional on the space of possible Reeb vector fields. 
An interesting corollary of this extremal problem is that
the normalized volumes of Sasaki-Einstein manifolds, and hence the $a$ central charges of the dual SCFTs in $d=4$, 
are necessarily algebraic numbers.

There is also a parallel story that involves AdS$_4\times SE_7$ solutions of eleven-dimensional
 supergravity, where $SE_7$ is a $d=7$ Sasaki-Einstein manifold, and the four-form flux is purely electric. In this case the dual field theories are three-dimensional ${\cal N}=2$ SCFTs that arise from membranes lying at the apex of the 
Calabi-Yau cone with link $SE_7$. These SCFTs again have an abelian R-symmetry, which is now determined by extremizing the supersymmetric free energy, namely minus the logarithm of a supersymmetric partition function \cite{Jafferis:2010un}. 
The volume minimization of \cite{Martelli:2005tp,Martelli:2006yb}, which is valid for Sasaki-Einstein manifolds
of arbitrary odd dimension greater than three, also provides a geometric description of this field theory
variational problem.

We now return to the general class of supersymmetric AdS$_3\times Y_7$ solutions of type IIB supergravity that will be the main focus of this paper. This class of solutions was first discussed in \cite{Kim:2005ez} and the geometry of $Y_7$ was further elucidated in \cite{Gauntlett:2007ts}. As one might expect, the compact manifolds $Y_7$ share several similarities with Sasaki-Einstein manifolds. For example, they also have a non-vanishing R-symmetry Killing vector field and the cone over $Y_7$ is again complex. However, there
are some crucial differences; for example, we will show here that the metric on $Y_7$ can never be Sasakian 
(in particular the R-symmetry Killing vector is never a Reeb vector). 
Furthermore, ensuring that the five-form flux is suitably quantized is more involved for the 
AdS$_3\times Y_7$ solutions, as compared to their AdS$_5\times SE_5$ cousins.

In presenting the geometric dual of $c$-extremization for the AdS$_3\times Y_7$ solutions of \cite{Kim:2005ez}, we need to find an 
appropriate way of going off-shell and, {\it a priori}, there is not a canonical procedure to do this. The approach we pursue here is to consider a specific class of geometries on a complex cone and admitting certain Killing spinors of the type discussed in \cite{Gauntlett:2007ts}, but relaxing the equation of motion for the five-form. A key point is that we also need to impose a natural topological constraint in order to ensure that the five-form flux is properly quantized.
With this set-up, the link of the cone still has an R-symmetry vector which,
moreover, foliates the link with a transversely conformal K\"ahler metric. A 
main result of this paper is to
show that the central charge $c_R$ of the dual SCFT can be obtained by extremizing a specific functional that depends on the space of R-symmetry vectors as well as the basic cohomology class of the transverse K\"ahler form.

As we shall see, these complex cone geometries and the related extremal problem can be formulated for all the geometries
$Y_{2n+1}$ with $n\ge 3$ introduced in \cite{Gauntlett:2007ts}. In particular, the results are also applicable to a class of supersymmetric AdS$_2\times Y_9$ solutions of eleven-dimensional supergravity, with only electric four-form flux, introduced in \cite{Kim:2006qu}, which are holographically dual to superconformal quantum mechanics with two supercharges. 
We will show that a naturally defined two-dimensional Newton constant, $G_2$, is obtained from the extremization
principle. Although determining the precise holographic dictionary for AdS$_2$ is still a work in progress (for some recent discussion and references, see {\it e.g.} \cite{Bena:2018bbd}), one can expect that $1/(4G_2)$ determines the logarithm of the partition function of the dual superconformal quantum mechanics.

For a certain sub-class of AdS$_2\times Y_9$ solutions we can also make a connection with recent work on microstate counting of supersymmetric AdS$_4$ black holes 
\cite{Benini:2015eyy,Hosseini:2016tor,Benini:2016rke,Cabo-Bizet:2017jsl,Azzurli:2017kxo,Liu:2017vbl}. Specifically, we can consider the class of solutions 
of the form AdS$_2\times \Sigma_g\times SE_7$, where $\Sigma_g$ is a Riemann surface with genus $g>1$, $SE_7$ is
a Sasaki-Einstein manifold and the $SE_7$ is fibred over the $\Sigma_g$ just in the direction of 
the Reeb vector of the $SE_7$. These solutions arise as the near horizon limit of black holes that asymptotically approach AdS$_4\times SE_7$
and hence have a clear dual interpretation. Specifically, after compactifying $D=11$ supergravity on $SE_7$ one gets an $\mathcal{N}=2$
SCFT in $d=3$. One can then further compactify this on the Riemann surface $\Sigma_g$, with the addition
of R--symmetry magnetic flux (only) on $\Sigma_g$; this is the so-called ``universal twist". It has been shown in \cite{Azzurli:2017kxo}
that the entropy of the black holes, $S_{BH}$, is precisely equal to the logarithm of the 
twisted topological index
\cite{Benini:2015noa,Benini:2016hjo,Closset:2016arn} and furthermore, that this is equal to minus the on-shell action
of the full AdS$_4$ black hole solution. A point that we make here is that the entropy is simply related to $G_2$ via
$S_{BH}=1/(4G_2)$, and hence, for this class of black hole solutions, we can obtain $S_{BH}$, as well as the twisted
topological index, via a novel variational principle.

In section \ref{sec:general} of this paper we will present several new results concerning the general class of odd-dimensional ``GK geometries"
$Y_{2n+1}$ of \cite{Gauntlett:2007ts}. In particular, after significant preparation in earlier subsections,
the extremal problem is presented in section \ref{sec:c}.
We will then restrict our attention,
in the remainder of the paper, to the special class of geometries with $Y_7=T^2\times Y_5$,
for which we make additional progress. These are 
associated with AdS$_3\times T^2\times Y_5$ solutions of type IIB supergravity, our main focus, or 
AdS$_2\times T^4\times Y_5$ solutions of $D=11$ supergravity. 
When $Y_7=T^2\times Y_5$ we can show that
the central charge $c_R$ of the AdS$_3\times T^2\times Y_5$ solutions
can be obtained by resolving a complex cone over $Y_5$ and then using localization, somewhat analogous to what was achieved in the Sasaki-Einstein case \cite{Martelli:2005tp,Martelli:2006yb}. In addition, we will also prove
an interesting obstruction theorem: if the cone over $Y_5$ is Calabi-Yau, then a supersymmetric solution of the form
AdS$_3\times T^2\times Y_5$ with the given complex structure on the cone does not exist.

It is interesting to point out that several infinite classes of explicit AdS$_3\times Y_7$ solutions of the type we are considering (and also AdS$_2\times Y_9$ solutions) have been known for some time \cite{Gauntlett:2006af,Gauntlett:2006qw,Gauntlett:2006ns,Donos:2008ug}, but the dual field theories for most of them have not yet been identified. Since the type IIB solutions have only non-trivial five-form flux, it is natural to consider them as arising from configurations involving a large number of D3-branes wrapping a complex submanifold inside a Calabi-Yau four-fold. For certain classes of solutions 
where $Y_7$ is the total space of a $Y_5$ fibration over $\Sigma_g$,
where $\Sigma_g$ is a Riemann surface, one can also anticipate that the dual $d=2$ field theories might arise as the low energy limit of some four-dimensional  ``parent'' SCFTs that have been appropriately compactified on $\Sigma_g$. Indeed some interesting progress in this direction
has been made in \cite{Benini:2013cda,Benini:2015bwz}. 

Here we will critically re-examine the possibility that specific examples of AdS$_3\times T^2\times Y_5$ solutions
have $d=2$ SCFTs duals which arise from compactifications of certain
$d=4$ quiver gauge theories dual to specific AdS$_5\times SE_5$ solutions, as discussed in \cite{Benini:2015bwz}.
More precisely, the idea is to consider the quiver gauge theory compactified on $T^2$ with vanishing flavour flux and non-vanishing baryonic flux on the $T^2$. In \cite{Benini:2015bwz} it was suggested that this possibility is realized for an 
explicit class of solutions first found in \cite{Donos:2008ug}, which we label here as
AdS$_3\times T^2\times \Ypq$, and the quiver gauge theories \cite{Benvenuti:2004dy} that are dual to the AdS$_5\times Y^{p,q}$ solutions \cite{Gauntlett:2004yd}.
Although some evidence for this was provided in \cite{Benini:2015bwz}, including what seemed to be a remarkable
matching of central charges as functions of $p,q$,
here we will show, surprisingly, that this possibility is in fact not realized. It remains an interesting open problem to identify the SCFTs dual to the AdS$_3\times T^2\times \Ypq$ solutions as well as to determine the fate of the $d=4$ $Y^{p,q}$ quiver gauge theories after they have been compactified on $T^2$, with only baryon flux on the $T^2$.

This analysis will be carried out in section \ref{sec:examples}, where we will also discuss some other examples of AdS$_3\times T^2\times Y_5$ solutions. In particular, we carry out the regularity analysis and also flux quantization for the local explicit solutions presented in
\cite{Donos:2008hd}, obtaining a new class of solutions which we label AdS$_3\times T^2\times \Labc$. The $\Labc$ metrics share
several similarities with the $L^{a,b,c}$ Sasaki-Einstein metrics of \cite{Cvetic:2005ft}. However, as in the previous paragraph, there is no direct connection between the $L^{a,b,c}$ quiver gauge theory \cite{Benvenuti:2005ja,Butti:2005sw,Franco:2005sm} compactified on $T^2$ and the 
AdS$_3\times T^2\times \Labc$ solutions.

The plan of the rest of the paper is as follows.
In section \ref{sec:general} we set up the general geometric formalism, reviewing and extending the results of 
\cite{Gauntlett:2007ts}, and then discuss the variational problem. In section \ref{sec:examplesT2} we specialize to $Y_7=T^2\times Y_5$ and show, in particular, that the central charge can be obtained via localization. Section~\ref{sec:examples} discusses several specific classes of solutions with
$Y_7=T^2\times Y_5$, including $Y_5=\Ypq$ and $\Labc$. Section~\ref{sec:examples} also contains an 
analysis of quiver gauge theories, dual to certain Sasaki-Einstein spaces, after they have been reduced on $T^2$.
We conclude in section \ref{sec:discuss} with some discussion of our results and open questions.


\section{General formalism}\label{sec:general}

In this section we study a general class of supersymmetric AdS$_3\times Y_7$ solutions of type IIB supergravity
and AdS$_2\times Y_9$ solutions of $D=11$ supergravity.
The geometric structure of $Y_7$ and $Y_9$, which share many similarities,
was first described in \cite{Kim:2005ez} and \cite{Kim:2006qu}, respectively, and then further clarified and also generalized to
higher odd dimensions, $Y_{2n+1}$, in \cite{Gauntlett:2007ts}. Here we will summarize some of these results and also significantly extend them. We introduce the relevant geometry in section
\ref{sec:background} and then discuss the complex geometry on the cones over $Y_{2n+1}$ in section  \ref{sec:cone}. In sections~\ref{sec:action} and \ref{sec:flux} we show 
that both the action and flux quantization conditions for a natural class of off-shell geometries may be written in terms of certain global, 
topological data. A supersymmetric solution is necessarily a critical point of this action. In section \ref{sec:c} we 
summarize this extremal problem, and show for $n=3$ that at a critical point the action coincides with the supergravity formula 
for the central charge of the dual $(0,2)$ SCFT. For $n=4$ the critical point determines the value of the two-dimensional Newton constant, $G_2$, which is related to the partition function of the dual quantum mechanics
and also gives, for a subset of AdS$_2$ solutions, the entropy of a class of AdS$_4$ black hole solutions.

\subsection{Geometric backgrounds}\label{sec:background}

We consider AdS$_3$ solutions of type IIB supergravity where the ten-dimensional metric and Ramond-Ramond 
self-dual five-form $F_5$ are given by\footnote{This is in agreement with the type IIB conventions in appendix A of \cite{Donos:2008ug}, and we note that we have a slightly  different expression here for the five-form than in equation (3.8) of
\cite{Gauntlett:2007ts}. We also note that the two-form $F_2$ appearing in appendix A of \cite{Donos:2008ug} (when $n=1$)
for type IIB/$D=11$ supergravity is minus/plus that appearing in \cite{Gauntlett:2007ts}.} 
\bea
\diff s^2_{10} &=& L^2 \ex^{-B/2}\left(\diff s^2_{\mathrm{AdS}_3} + \diff s^2_{7}\right)~,\nn\\
F_5 &=& -L^4\left(\vol_{\mathrm{AdS}_3}\wedge F + *_7 F\right)~.\label{ansatz}
\eea
We also consider AdS$_2$ solutions of $D=11$ supergravity where the eleven-dimensional metric and four-form $G$ are given by
\bea
\diff s^2_{11} &=& L^2 \ex^{-2B/3}\left(\diff s^2_{\mathrm{AdS}_2} + \diff s^2_{9}\right)~,\nn\\
G &=& L^3\vol_{\mathrm{AdS}_2}\wedge F ~.\label{ansatzd11}
\eea
In these expressions $L$ is an overall dimensionful length scale, with $\diff s^2_{\mathrm{AdS}_3}$ and
$\diff s^2_{\mathrm{AdS}_2}$ being the metrics on a 
unit radius AdS$_3$ and AdS$_2$, respectively, with corresponding volume forms $\vol_{\mathrm{AdS}_3}$ and $\vol_{\mathrm{AdS}_2}$. In each case the warp factor $\BB$ is a function on the smooth, compact Riemannian internal space $(Y_7, \diff s^2_7)$ or $(Y_9, \diff s^2_9)$ while $F$ is a closed two-form on $Y_7$ or $Y_9$ with, in the former case, Hodge dual $*_7 F$. 

The ansatz (\ref{ansatz}) is the most general 
AdS$_3$ background one can write down that has only metric and five-form turned on.
Due to the presence of only five-form flux, they are in some sense D3-brane backgrounds. 
In particular, if one wraps a large number of D3-branes over a Riemann surface $\Sigma_g$, assuming the resulting 
two-dimensional low-energy theory flows to a CFT in the IR one would expect this to have a holographic dual 
of the form (\ref{ansatz}).
The
ansatz \eqref{ansatz} is the natural analogue of the Freund-Rubin 
AdS$_5\times SE_5$  ansatz, which describes holographic duals of unwrapped D3-brane worldvolume theories in flat space sitting at the apex of the cone over $SE_5$.
In the latter case, supersymmetry implies that $SE_5$ is a Sasaki-Einstein manifold \cite{Acharya:1998db} and is realized when
the cone over $SE_5$ is Calabi-Yau.
On the other hand, supersymmetric AdS$_3\times Y_7$ solutions of the form (\ref{ansatz}), which are dual to SCFTs preserving\footnote{If one wants to obtain AdS$_3$ solutions that preserve $(0,1)$ supersymmetry, one should allow for three-form flux. A concrete brane realization is D3-branes wrapping a holomorphic curve in a Calabi-Yau four-fold, combined with five-branes wrapping a SLAG four-cycle.} 
$(0,2)$ supersymmetry \cite{Kim:2005ez}, can arise when D3-branes wrap holomorphic cycles in Calabi-Yau four-folds.
As we will review below, $Y_7$ in the supersymmetric AdS$_3\times Y_7$ solutions
share a number of features with Sasaki-Einstein geometry \cite{Gauntlett:2007ts}.

Similar comments apply to the ansatz \eqref{ansatzd11}. It is the most general 
AdS$_2$ background with purely electric four-form flux and is associated with M2-brane backgrounds. 
If one wraps a large number of M2-branes over a Riemann surface $\Sigma_g$, assuming the resulting 
one-dimensional low-energy theory flows to a conformal quantum mechanics
in the IR one would expect this to have a holographic dual 
of the form \eqref{ansatzd11}. It is a natural analogue of the Freund-Rubin AdS$_4\times SE_7$ ansatz.
If we demand supersymmetry, $SE_7$ is again a Sasaki-Einstein manifold, while supersymmetric AdS$_2\times Y_9$ solutions
are dual to supersymmetric quantum mechanics preserving 2 supercharges\footnote{If one wants AdS$_2$ solutions dual to quantum mechanics preserving one supercharge one should add magnetic four-form flux.
A concrete brane realization is M5-branes wrapping a SLAG five-cycle inside a Calabi-Yau five-fold, with the option of
also having M2-branes wrapping a holomorphic curve \cite{Gauntlett:1998vk}.} and can arise from membranes wrapping holomorphic curves in a Calabi-Yau five-fold. It is also worth noting that AdS$_2\times Y_9$ (supersymmetric) solutions with $Y_9=T^2\times Y_7$
can be dimensionally reduced to type IIA and then T-dualized to type IIB, and we find precisely the (supersymmetric) 
AdS$_3\times Y_7$ solutions in \eqref{ansatz} \cite{Gauntlett:2006ns}.

Substituting the ansatz (\ref{ansatz}) and \eqref{ansatzd11} into the type IIB and $D=11$ equations of motion, respectively,
gives corresponding equations of motion for the metric 
$\diff s^2_{2n+1}$ on $Y_{2n+1}$, function $B$, and local one-form $A$ with curvature $F\equiv \diff A$.
Here $n=3$ is the type IIB AdS$_3$ case and $n=4$ is the $D=11$ AdS$_2$ case.
As shown in \cite{Gauntlett:2007ts}, these equations of motion can in turn be derived from the action
\bea\label{action}
S \ = \ \int_{Y_{2n+1}} \ex^{(1-n)B}\left[R_{2n+1} - \frac{2n}{(n-2)^2}+\frac{n(2n-3)}{2}(\diff B)^2 + \frac{1}{4}\ex^{2B}F^2\right]\vol_{2n+1}~.
\eea
Here we have written an action in general dimension $2n+1$, with Riemannian manifold $Y_{2n+1}$ having Ricci scalar $R_{2n+1}$ and 
Riemannian volume form $\vol_{2n+1}$, and $F^2\equiv F_{ab}F^{ab}$. 

Demanding that the AdS$_3\times Y_7$ and AdS$_2\times Y_9$ backgrounds preserve
supersymmetry leads to the existence of specific Killing spinors on $Y_{2n+1}$, for $n=3,4$,
and this implies additional geometric structure. It was shown in \cite{Gauntlett:2007ts} that
these properties can also be generalized to $n>4$, so we continue with arbitrary $n\ge 3$.
Supersymmetry, by which we now mean the existence of the Killing spinors given in \cite{Gauntlett:2007ts},
implies that the metric on $Y_{2n+1}$ is equipped with a 
unit norm Killing vector field $\xi$, which we call the \emph{R-symmetry vector field}. In local coordinates we write 
\bea\label{Reeb}
\xi &=& \frac{1}{\cc}\partial_\z~, \qquad \mbox{where} \quad     \cc \ \equiv \ \frac{1}{2}(n-2)~.
\eea
Since $\xi$ is nowhere zero, it defines a foliation $\mathcal{F}_\xi$ of $Y_{2n+1}$. 
The metric takes the form
\bea\label{metric}
\diff s^2_{2n+1} &=& \cc^2(\diff \z + P)^2 + \ex^{\BB} \diff s^2~,
\eea
where $\diff s^2$ is a K\"ahler metric, transverse to the foliation $\mathcal{F}_\xi$. In fact this K\"ahler metric 
in real dimension $2n$ determines all of the remaining fields. The function $\BB$ is fixed via
\bea 
\ex^{\BB} &=& \frac{\cc^2}{2}R~,\label{fixB}
\eea
where $R$ is the Ricci scalar of the transverse K\"ahler metric. 
Notice here that we need positive scalar curvature, $R>0$, in order that the metric (\ref{metric}) is well-defined and positive definite.\footnote{When $R<0$ one can obtain \cite{Gauntlett:2006ns}, after a double wick rotation, supersymmetric solutions of type IIB and
$D=11$ with $S^3$ and $S^2$ factors, respectively, that are also of interest, but we will not discuss them further here.}
 The local one-form $P$ in (\ref{metric}) is the Ricci one-form of the transverse K\"ahler metric, so that 
$\diff P = \rho$ is the Ricci two-form. Finally, the closed two-form  is given by
 \bea\label{fixF}
 F &=& -\frac{1}{\cc}J + \cc\, \diff\left[\ex^{-B}(\diff \z + P)\right]~,
 \eea
 where $J$ is the transverse K\"ahler form. 

The above geometric conditions on $Y_{2n+1}$ are equivalent to the existence of a non-trivial solution to the Killing spinor 
equations given in \cite{Gauntlett:2007ts}. Moreover, imposing also the equation of motion for the two-form,
\bea\label{Maxwell}
\diff \left[\ex^{(3-n)\BB}*_{2n+1} F\right] &=& 0~,
\eea
the supersymmetric backgrounds given by (\ref{metric})--(\ref{fixF}) automatically obey the equations of motion 
derived from the action (\ref{action})  \cite{Gauntlett:2007ts}. On the other hand, it is straightforward to show that 
(\ref{Maxwell}) 
is equivalent to the PDE
\bea\label{boxR}
\Box R &=& \frac{1}{2}R^2 - R_{ij}R^{ij}~,
\eea
where $R_{ij}$ denotes the transverse Ricci tensor, and everything in (\ref{boxR}) is computed using the transverse K\"ahler metric. We shall refer to equation (\ref{boxR}) as the equation of motion
in what follows. Moreover, to be clear, we refer to geometries satisfying (\ref{metric})--(\ref{fixF}) as 
\emph{supersymmetric geometries} (since they solve the Killing spinor equations), and 
if the equation of motion (\ref{boxR}) also holds these are then \emph{supersymmetric solutions}. 
Thus, the supersymmetric geometries are ``off-shell'' in a precise sense which we utilize below.

The above supersymmetric geometry is at first glance very similar to Sasakian geometry. 
In both cases there is a  unit norm Killing vector field $\xi$, which foliates the manifold $Y_{2n+1}$ with 
a transversely conformal K\"ahler metric. However, in the present setting the dual one-form
\bea
\eta &\equiv & \cc (\diff\z +P)~,
\eea 
satisfies $\diff\eta = \cc \rho$, where recall that $\rho$ is the transverse Ricci form. 
On the other hand, in Sasakian geometry instead $\diff\eta$ is proportional to the transverse \emph{K\"ahler} form 
$J$, implying that  $Y_{2n+1}$ is a contact manifold. In fact in section \ref{sec:flux} we will see that 
Sasakian manifolds never solve the equation of motion (\ref{boxR}), and in this sense the supersymmetric geometry 
we have is orthogonal to Sasakian geometry. 

\subsection{Complex geometry}\label{sec:cone}

Another feature in common with Sasakian geometry is that the real cone over 
$Y_{2n+1}$ is a complex 
cone  \cite{Gauntlett:2007ts}. Here one introduces the $(2n+2)$-dimensional cone $C(Y_{2n+1})\equiv \R_{>0}\times Y_{2n+1}$, 
equipped with the conical metric
\bea
\diff s^2_{2n+2} &=& \diff r^2 + r^2 \diff s^2_{2n+1}~,
\eea
where $r>0$ is a coordinate on $\R_{>0}$. Notice that we exclude the tip of the cone $r=0$, so that 
$C(Y_{2n+1})$ is a smooth manifold.
There is a natural compatible $SU(n+1)$ structure on this cone, with 
fundamental two-form $\mathcal{J}$ and holomorphic volume form $\Omega_{(n+1,0)}$ given by
\bea\label{SUstructure}
\mathcal{J} &=& -\cc\, r\diff r\wedge (\diff \z + P) + r^2 \ex^\BB J~,\nonumber\\
\Omega_{(n+1,0)}&=& \ex^{\ii \z} \left(\ex^{\BB/2}r\right)^n \left[\diff r - \ii r \cc (\diff \z + P)\right]\wedge \Omega~.
\eea
Here $\Omega$ is a local $(n,0)$-form for the transverse K\"ahler metric, satisfying 
\bea\label{dOmega}
\diff \Omega &=& \ii P \wedge \Omega~,
\eea
where recall that $\diff P = \rho$. 
The real two-form $\mathcal{J}$ is not closed, nor conformally closed, and so there is no (natural) symplectic 
structure on $C(Y_{2n+1})$. However, $\Omega_{(n+1,0)}$ is both globally defined and conformally closed:
\bea\label{Psidef}
\diff\Psi &=& 0~, \qquad \mbox{where}\ \, \Psi \ \equiv \ \ex^{-n\BB/2}r^{-\frac{n(n-1)}{(n-2)}}\Omega_{(n+1,0)}~.
\eea
This implies  that $C(Y_{2n+1})$ has an integrable complex structure, with zero first Chern class. 
We discuss this in more detail below. We note that $\Omega_{(n+1,0)}$, or equivalently $\Psi$, has charge $1/\cc$ under the R-symmetry 
vector field $\xi=\frac{1}{\cc}\partial_\z$:
\bea\label{chargeOmega}
\mathcal{L}_\xi \Psi &=& \frac{\ii}{\cc} \Psi~.
\eea
This implies that $\xi$ is a holomorphic vector field. It pairs with the radial vector field $r\partial_r = -\mathcal{I}(\xi)$ under the 
complex structure tensor\footnote{We use conventions where $\mathcal{I}$ is obtained by raising an index on {\it minus} the K\"ahler two-form.}  $\mathcal{I}$. The foliation generated by $\xi$ is then also \emph{transversely holomorphic}, 
meaning that there are local coordinate patches $\R\times \C^n$, with corresponding local coordinates $\z,z_1,\ldots,z_n$, where $(z_1,\ldots,z_n)$ are complex coordinates on $\C^n$. In particular the latter transform holomorphically between coordinate patches. For further details on this, and some of the other material in this subsection, see section 1 of \cite{Sparks:2010sn} 
and references therein.

 It will be convenient for what follows to introduce the \emph{basic cohomology} of the foliation $\mathcal{F}_\xi$. 
A form $\alpha$ on $Y_{2n+1}$ is called \emph{basic} if 
\bea\label{basic}
\xi\lrcorner \alpha & =&  0~, \qquad \mathcal{L}_\xi \alpha \ = \ 0~.
\eea
The basic cohomology $H^*_{{B}}(\mathcal{F}_\xi)$ is by definition simply the cohomology defined by the exterior derivative 
$\diff$ restricted to basic forms on $Y_{2n+1}$. For example, the transverse K\"ahler form $J$ is basic and closed, and so defines a class 
$[J]\in H^2_{{B}}(\mathcal{F}_\xi)$. The one-form $\eta$ is not basic, but its exterior derivative $\diff\eta = \cc \rho$ is, and 
in particular $[\rho]\in H^2_{{B}}(\mathcal{F}_\xi)$ also defines a basic cohomology class. Since basic cohomology classes 
are by definition represented by closed (basic) forms on $Y_{2n+1}$, there is a natural map 
\bea\label{forget}
H^*_{{B}}(\mathcal{F}_\xi) \ \rightarrow \ H^*(Y_{2n+1},\R)~.\eea 
Note that
$[\rho]$ lies in the kernel of this map, since $\rho = \frac{1}{\cc}\diff\eta$. Since the foliation is also transversely holomorphic, we may 
further grade the basic cohomology by Hodge type. In particular the transverse Ricci form $\rho$ 
has Hodge type $(1,1)$, and $[\rho/2\pi]$ represents the basic cohomology class 
$c_1^B\in H^{1,1}_B(\mathcal{F}_\xi)$. This is independent of the choice of transverse K\"ahler metric, 
and depends only on the transversely holomorphic foliation $\mathcal{F}_\xi$. 
Similarly, the K\"ahler form is Hodge type $(1,1)$, so we may more accurately write 
$[J]\in H^{1,1}_{{B}}(\mathcal{F}_\xi)$. 

Consider now a general class of supersymmetric geometries of the form (\ref{metric})--(\ref{fixF}). 
We fix both the choice of manifold $Y_{2n+1}$ and its complex cone $C(Y_{2n+1})$. 
However, we are still free to  choose a nowhere zero holomorphic vector field $\xi$, 
together with a transverse K\"ahler metric. In fact any choice of such vector field and K\"ahler metric 
defines a supersymmetric geometry via  (\ref{metric})--(\ref{fixF}), provided the scalar curvature $R>0$. 
This class of off-shell geometries may be described as follows. 
We fix a choice of complex 
manifold $C(Y_{2n+1})\cong \R_{>0}\times Y_{2n+1}$, with closed 
holomorphic volume form $\Psi$ proportional to $\Omega_{(n+1,0)}$ (\ref{Psidef}). Since $\Psi$ has fixed charge $1/\cc$ 
under $\xi$, we may write a general choice of R-symmetry vector field as 
\bea\label{trialR}
\xi &=& \sum_{i=1}^s b_i\partial_{\varphi_i} \ = \ \xi_0 + \sum_{i=2}^{s} c_i \partial_{\varphi_i}~.
\eea
Here in the first expression $\partial_{\varphi_i}$, $i=1,\ldots,s\geq 1$, are real holomorphic
vector fields generating a $U(1)^{s}$ action on $C(Y_{2n+1})$. We choose this basis so that 
the holomorphic volume form has unit charge under $\partial_{\varphi_1}$, but is 
uncharged under $\partial_{\varphi_i}$, $i=2,\ldots,s$.
Physically, the latter will be dual to non-R global symmetries. Notice this fixes the coefficient $b_1=1/\cc$.
For some considerations also the second expression in (\ref{trialR}) will be useful. 
Here 
$\xi_0$ is a particular fiducial choice of R-symmetry vector, meaning that $\Psi$ has 
charge $1/\cc$ under $\xi_0$. 
 We then have a family of R-symmetry 
vector fields in (\ref{trialR}), parametrized either by $(b_1=1/\cc,b_2\ldots,b_s)$, or equivalently by 
 $(c_1,\ldots,c_{s-1})$. 

Note that contracting the 
complex holomorphic vector field $\xi -\ii \mathcal{I}(\xi)$ into $\Psi$ gives a global 
 $(n,0)$-form that is transverse to the corresponding foliation $\mathcal{F}_\xi$. This is proportional to the form 
$\ex^{\ii \z}\, \Omega$, where the phase-dependence is fixed by the fixed charge 
of $\Psi$ under $\xi=\frac{1}{\cc}\partial_\z$, and $\Omega$ is a local transverse $(n,0)$-form. 
In fact neither factor in $\ex^{\ii \z}\, \Omega$ is globally defined separately, with the transition functions between patches 
for each factor precisely cancelling. Indeed,  $\Omega$ is a section of the basic canonical line bundle
$\Lambda^{n,0}_B$, whose basic first Chern class is $-c_1^B\in H^{1,1}_B(\mathcal{F}_\xi)$ -- see equation 
(\ref{dOmega}). 
Since $\diff\z + P$ is a global one-form on $Y_{2n+1}$, where recall that 
$[\rho/2\pi]=c_1^B\in H^{1,1}_B(\mathcal{F}_\xi)$, 
this means that $\ex^{\ii \z}$ is a section of the dual line bundle. 
Having chosen an R-symmetry vector field and hence transversely holomorphic 
foliation, we are then free to pick a compatible transverse K\"ahler metric. 
In particular, this involves specifying a choice of the basic 
K\"ahler class $[J]\in H^{2}_B(\mathcal{F}_\xi)$.

Finally, as in Sasakian geometry we may classify these geometries according to whether the orbits of 
$\xi$ all close or not. If $\xi$ has a non-closed orbit we call the structure \emph{irregular}, while 
if all orbits close, and are hence circles, then we call the structure \emph{quasi-regular}. In the latter 
case $\xi$ defines a $U(1)$ action on $Y_{2n+1}$. If this action is free then the structure is called \emph{regular}, 
and in that case $Y_{2n+1}$ is the total space of a circle bundle over a compact K\"ahler manifold 
$(V,J)$. In this case $H^*_{{B}}(\mathcal{F}_\xi)\cong H^*(V,\R)$ is naturally isomorphic to the 
cohomology of the base $V$, and moreover 
 $[\rho/2\pi ]=c_1\in H^2(V,\Z)$ represents an integral cohomology class, the first Chern class.
The largest positive integer $I_V$ such that $c_1/I_V \in H^2(V,\Z)$ is called the \emph{Fano index} 
of $V$. In general the coordinate $\z$ may then have period $2\pi I_V/\mm$, where $\mm$ is a positive integer that divides $I_V$.
This latter condition is required in order that the Killing spinors, or equivalently the holomorphic volume form 
$\Omega_{(n+1,0)}$ in (\ref{SUstructure}), are appropriately single-valued.
Then $Y_{2n+1}$ is 
the total space of the circle bundle over $V$ associated to the line bundle $K^{\mm/I_V}$, where 
$K=\Lambda^{n,0}$ is the canonical line bundle of $V$. In fact all of these statements go through also in the quasi-regular case, when $V$ is an orbifold, provided one replaces 
$H^2(V,\Z)$ by the appropriate notion of orbifold cohomology group $H^2_{\mathrm{orb}}(V,\Z)$. 

\subsection{The action}\label{sec:action}

Consider the general class of supersymmetric geometries of the form (\ref{metric})--(\ref{fixF}) described 
in the previous subsections. In particular we have a choice of nowhere zero holomorphic vector field $\xi$, 
of the form (\ref{trialR}), and having fixed $\xi$ then a choice of compatible transverse K\"ahler metric. 
One can then restrict the original action (\ref{action}) to this off-shell class of supersymmetric geometries. 
They are off-shell because we do not (yet) impose the equation of motion (\ref{boxR}). 
A straightforward computation leads to the remarkably simple formula
\bea\label{Ssusy}
\Ssusy &=& \int_{Y_{2n+1}} \eta\wedge \rho \wedge \frac{J^{n-1}}{(n-1)!}~.
\eea
Here we have dropped a total derivative term using Stokes' Theorem and the fact that $Y_{2n+1}$ is compact without boundary.
For future use, we note that this can also be written in the form\footnote{
We can also write
$\frac{(n-1)}{2}\Ssusy=\frac{1}{4}\int_{Y_{2n+1}}\ex^{(3-n)B}F^2\vol_{2n+1}$, where the right hand side is the last term appearing in 
\eqref{action}. In addition
$\Ssusy = c\int_{Y_{2n+1}} [(\nabla B)^2+\frac{1}{2}c^2\ex^{-B}\rho^2]\vol_{2n}$, where
in this particular expression indices are raised using the $2n$-dimensional transverse K\"ahler metric $\dd s^2$ appearing in
\eqref{metric} and $\vol_{2n}$ is the corresponding volume form.}
\bea\label{Ssusy2}
\Ssusy &=& \frac{1}{c^2}\int_{Y_{2n+1}} \ex^{(1-n)B}\vol_{2n+1}~.
\eea
Notice that $\Ssusy$ must be positive.

The action (\ref{Ssusy}) is a functional of both the choice of R-symmetry vector field $\xi$, and also the transverse K\"ahler metric. 
However, an immediate observation is that $\Ssusy$ only depends on this K\"ahler metric via its 
 K\"ahler class $[J]\in H^2_B(\mathcal{F}_\xi)$. To be precise, suppose that $J_1$, $J_2$ are two K\"ahler forms related via
\bea\label{samebasic}
J_2 &=& J_1 + \diff \alpha~,
\eea
where $\alpha$ is a basic one-form on $Y_{2n+1}$, thus satisfying (\ref{basic}).
The corresponding 
Ricci forms are then similarly related by $\rho_2=\rho_1+\diff\beta$, where $\beta$ is also a basic one-form. 
An integration by parts then shows that the actions for $J_1$ and $J_2$ are equal.
To see this, note that if $\Phi$ is any closed form with $\xi\lrcorner\Phi=0$, then
\bea
\int_{Y_{2n+1}}\eta\wedge \diff\alpha\wedge \Phi \ =\ \int_{Y_{2n+1}}\diff\eta\wedge\alpha\wedge\Phi \ = \ 0~.
\eea
The first equality follows from Stokes' Theorem (and integration by parts), while the second follows since 
the integrand has zero contraction with $\xi$ and is hence identically zero.
We have thus shown that $\Ssusy$ depends on just the basic classes $[J]$ and $[\rho]$ both in
$H^{1,1}_B(\mathcal{F}_\xi)$. Since $[\rho]\in H^{1,1}_B(\mathcal{F}_\xi)$ is an invariant of the 
transversely holomorphic foliation $\mathcal{F}_\xi$, independent of the choice of 
K\"ahler metric one uses to compute $\rho$,
we deduce that $\Ssusy$ depends on
$\xi$ and this basic K\"ahler class $[J]\in H^{1,1}_B(\mathcal{F}_\xi)$, and we have
\begin{align}
\Ssusy \ = \ \Ssusy(\xi; [J])~.
\end{align}

We conclude this subsection by noting that for quasi-regular structures the action (\ref{Ssusy}) may be written  as
\bea\label{Ssusyquasi}
\Ssusy &=& \frac{(2\pi)^2\cc I_V}{\mm}\int_V c_1\wedge \frac{J^{n-1}}{(n-1)!} \ = \ \frac{\pi \cc I_V}{\mm}\int_V R \, \frac{J^n}{n!}~.
\eea
Here the notation is the same as that at the end of section \ref{sec:cone}, with $I_V$ being the (orbifold) 
Fano index of $V=Y_{2n+1}/U(1)$.
For fixed quasi-regular vector field $\xi$, the middle expression in (\ref{Ssusyquasi}) is  more manifestly a function of the K\"ahler class $[J]\in H^2(V,\R)\cong H^2_{{B}}(\mathcal{F}_\xi)$. 

\subsection{Flux quantization}\label{sec:flux}

In order that solutions of the form (\ref{ansatz}) and \eqref{ansatzd11} define consistent backgrounds of
type IIB string theory and M-theory, respectively, we have to impose flux quantization. There are a number of subtleties that we need to discuss, including the fact that we want to impose a version of flux quantization for the off-shell
supersymmetric geometries.

We first discuss flux quantization for the type IIB supersymmetric AdS$_3$ solutions.
In order to define a consistent string theory background, 
the five-form must satisfy an appropriate Dirac quantization condition over all five-cycles $\Sigma_A\subset Y_7$,
where $A=1,\ldots, \mathrm{rank}\, H_5(Y_7,\Z)$ 
runs over an integral basis for the free part of $H_5(Y_7,\Z)$. 
Specifically, the condition is 
\bea
\frac{1}{(2\pi \ell_s)^4 g_s}\int_{\Sigma_A} F_5 &=& N_A \in \mathbb{Z}~,\label{quantization}
\eea
where $\ell_s$ is the dimensionful string length, and $g_s$ is the constant string coupling.
Now, for the supersymmetric geometries the relevant part of the five-form is the piece on $Y_7$ which is given by
\bea
F_5\mid_{Y_7} &=& \frac{L^4}{4}\left[(\diff \z+P) \wedge \rho\wedge J  + \frac{1}{2}*\diff R\right]~.\label{flux}
\eea
It is important to note that in the supersymmetric solutions $F_5\mid_{Y_7}$ is closed and hence
\eqref{quantization} only depends on the homology class of $\Sigma_A$. For supersymmetric geometries
$F_5\mid_{Y_7}$ is not closed and hence more care is required in imposing flux quantization, as we discuss below.

Before doing that, we next discuss flux quantization for the $D=11$ supersymmetric AdS$_2$ solutions, which has a few
distinctive subtleties. We first note that despite the fact that $G|_{Y_9}=0$ there is still a non-trivial condition concerning flux quantization of $G$. Specifically, from \cite{Witten:1996md} we require that the first Pontryagin class of $Y_9$,
$p_1(Y_9)$, is divisible by four, and so we will implicitly assume that this is the case.
Since $Y_9$ is a spin manifold we necessarily have  $p_1(Y_9)$ is divisible by two. 
The next subtlety concerns the quantization of the electric part of the four-form flux. The equation of motion for
the four-form in $D=11$ supergravity is given by 
 \begin{align}\label{fluxqnqc}
 \dd*_{11}G+\frac{1}{2}G\wedge G \ = \  0~.
 \end{align}
In general this requires that a suitably defined ``Page charge" is appropriately quantized on $Y_9$. However, for the ansatz
\eqref{ansatzd11} we have $G\wedge G=0$, so we just need to impose
\bea
\frac{1}{(2\pi \ell_p)^6}\int_{\Sigma_A} *_{11}G &=& N_A \in \mathbb{Z}~,\label{quantization11}
\eea
over all seven-cycles $\Sigma_A\subset Y_9$,
where $A=1,\ldots, \mathrm{rank}\, H_7(Y_9,\Z)$ runs over an integral basis for the free part of $H_7(Y_9,\Z)$, and 
$\ell_p$ denotes the eleven-dimensional Planck length. Now for the supersymmetric geometries we have,
on $Y_9$,
\begin{align}
*_{11}G \ = \ L^6\left[(\diff \z+P) \wedge \rho\wedge \frac{J^2}{2}  + \frac{1}{2}*\diff R\right]\,.
\label{flux11}
\end{align}
For the supersymmetric solutions $*_{11}G$ is closed, being equivalent to \eqref{boxR},
and hence \eqref{quantization11} only depends on the homology class of 
$\Sigma_A$. For supersymmetric geometries $*_{11}G$ is not closed and again care is required 
in imposing flux quantization. A final subtlety for the $D=11$ case is that the next order correction to the supergravity equations of motion gives rise to a contribution of $-\frac{(2\pi \ell_p)^6}{192}(p_1(Y_9)^2-4p_2(Y_9))$ 
appearing on the right hand side of \eqref{fluxqnqc}, where $p_2(Y_9)$
is the second Pontryagin form \cite{Vafa:1995fj,Duff:1995wd}. This term, 
which arises from anomaly considerations, is certainly important in properly imposing flux quantization, but it gives rise to corrections to the fluxes that are sub-leading in the large $N_A$ limit, and hence for simplicity we will not consider them further in this paper. 
We note that when $Y_9=T^2\times Y_7$, all of the above quantum subtleties involving Pontryagin classes are absent, and indeed 
in this case the solutions may be reduced and T-dualized to type IIB solutions where these corrections are indeed not present.\footnote{To see this, note first that for $M$ any spin manifold $p_1(M)/2$ is congruent modulo 2 to the fourth Stiefel-Whitney class
$w_4(M)$. On the other hand, for $M$  a spin seven-manifold $w_4(M)=0$ on dimensional grounds (for example, 
see \cite{Gukov:2001hf}), and $w_4(T^2\times Y_7)$ is simply a pull-back of $w_4(Y_7)=0$. Similarly, the curvature forms representing Pontryagin classes $p_1^2$ 
and $p_2$
of each of $T^2$ and $Y_7$ are both identically zero on dimensional grounds, which implies they vanish for the product.}

We are now in a position to discuss flux quantization for the off-shell supersymmetric geometries. We first observe that
for the AdS$_3\times Y_7$ and the AdS$_2\times Y_9$ cases, imposing closure of  
$F_5\mid_{Y_7} $ and $*_{11}G$ is equivalent to imposing the equation of motion (\ref{boxR}), which
would put us on-shell. However, we may instead impose the weaker condition that 
the \emph{integral} of (\ref{boxR}) over $Y_{2n+1}$ holds. A short computation reveals that (\ref{boxR}) may be rewritten as
\bea
\Box R &=& (J\wedge J)\lrcorner (\rho\wedge \rho)~.\label{boxRv2}
\eea
Since the integral of $\Box R$ over $Y_{2n+1}$, using the measure $\eta\wedge J^n/n!$, vanishes using Stokes' Theorem, a necessary condition 
to solve the equation of motion is 
\bea\label{constraint}
\int_{Y_{2n+1}} \eta \wedge \rho^2 \wedge \frac{J^{n-2}}{(n-2)!} &=& 0~.
\eea
Compare this expression to the supersymmetric action (\ref{Ssusy}): the left hand side is again only a function of the choice
of Killing vector  $\xi$ and basic K\"ahler class for a supersymmetric geometry. 
As we shall see in this subsection, (\ref{constraint}) is also a sufficient condition in order to be able to 
impose flux quantization when $n=3$ and $n=4$ for a supersymmetric geometry of type IIB or $D=11$ supergravity, respectively. 
We note in passing that Sasakian metrics with metric cones admitting a holomorphic $(n+1,0)$-form 
have $[\rho]$ being a positive multiple of $[J]$, and hence the left hand side of (\ref{constraint}) is then
a positive multiple of the Riemannian volume of $Y_{2n+1}$. Thus Sasakian metrics can never be used
for supersymmetric solutions of the type we are discussing.
In particular the R-symmetry Killing vector is never a Reeb vector. 

A sufficient topological condition to interpret (\ref{quantization}), \eqref{quantization11} for our supersymmetric geometries 
is that 
\bea\label{topcondition}
H^2(Y_{2n+1},\R) &  \cong & H^2_{{B}}(\mathcal{F}_\xi)/[\rho]~.
\eea 
Note that $\rho=\frac{1}{\cc}\diff\eta$ is automatically exact in $H^2(Y_{2n+1},\R)$,
as noted after equation (\ref{forget}). In fact the Gysin long exact sequence for the 
foliation, discussed for example in \cite{Boyer:2008era,Boyer:2004fc}, implies that $[\rho]$ spans the kernel of the map in (\ref{forget}). 
 The content of 
(\ref{topcondition}) is hence that the map in (\ref{forget}) is \emph{surjective} {\it i.e.}
\emph{all} closed two-form classes on $Y_{2n+1}$ can be represented by 
\emph{basic} closed two-forms. 
This holds in all examples of which we are aware. For example, it follows immediately from the Gysin long 
exact sequence for the foliation if $H^1_{{B}}(\mathcal{F}_\xi)=0$. In fact this latter condition 
does \emph{not} hold for the class of examples in section \ref{sec:examplesT2}, but as discussed in that section
(\ref{topcondition}) does hold. 

In the following discussion we briefly restrict to the quasi-regular case for simplicity, although notice that 
an irregular R-symmetry vector may be viewed as a limit of a sequence of quasi-regular vector fields 
(since irrational numbers are limits of sequences of rationals). Since the quantities of interest are all continuous,  the equations  we deduce
will hold also in the irregular case. For quasi-regular geometries we 
have  $H^2_{{B}}(\mathcal{F}_\xi)\cong H^2(V,\R)$, where $V=Y_{2n+1}/U(1)$ is 
the K\"ahler orbifold base. The dual homology statement to (\ref{topcondition}) implies that 
all $(2n-1)$-cycles in $H_{2n-1}(Y_{2n+1},\R)\cong H^2(Y_{2n+1},\R)$ may be represented as circle fibrations 
over $(2n-2)$-cycles in $V$. Because of this, the second $*\,  \diff R$ term in the flux (\ref{flux}) or \eqref{flux11} does not contribute to the integral, 
since it manifestly has zero contraction with $\xi$ which generates the circle action. This will be true for any submanifold representing 
$\Sigma_A$, that is everywhere tangent to $\xi$.
In this case  only the first term in  (\ref{flux}) contributes to the integral, leading to 
\begin{align}
\int_{\Sigma_A}\eta\wedge \rho\wedge \frac{J^{n-2}}{(n-2)!} &\ = \ \begin{cases} \ \displaystyle \frac{2(2\pi\ell_s)^4g_s}{L^4}\, N_A\,, \qquad n \, = \, 3~, \\
\ \displaystyle \frac{(2\pi \ell_p)^6}{L^6}\, N_A\,, \quad \ \qquad n \, = \, 4~.\end{cases}\label{quantize}
\end{align}
Again, compare the left hand side of (\ref{quantize}) to the key formulae (\ref{Ssusy}) and (\ref{constraint}). 
However, {\it a priori} the left hand side of (\ref{quantize}) is still not well-defined. 
We must require $\Sigma_A$ to be tangent to $\xi$, as already discussed, but consider two such submanifolds $\Sigma_A^{(1)}$, $\Sigma_A^{(2)}$
that represent the same homology class in $H_{2n-1}(Y_{2n+1},\R)$. In the quasi-regular case these are circle 
bundles over $(2n-2)$-dimensional subspaces $C_A^{(1)}$, $C_A^{(2)}$ of $V$ representing $(2n-2)$-cycles in $H_{2n-2}(V,\R)\cong H^2(V,\R)\cong H^2_{{B}}(\mathcal{F}_\xi)$. 
However, two such $(2n-2)$-cycles with Poincar\'e duals differing by a multiple of $[\rho]$ both lift to the \emph{same} $(2n-1)$-cycle, due to 
(\ref{topcondition}). Writing $C_A^{(2)}-C_A^{(1)}= \lambda [\rho]_{\mbox{\tiny{Poincar\'e dual}}} \in H_{2n-2}(V,\R)$, we then compute
\bea\label{consistent}
\int_{\Sigma_A^{(2)}}\eta\wedge \rho\wedge J^{n-2} - \int_{\Sigma_A^{(1)}}\eta\wedge \rho\wedge J^{n-2} \ = \ \lambda \int_{Y_{2n+1}}\eta\wedge \rho\wedge \rho \wedge J^{n-2}~.
\eea
Here $\lambda\in\R$ is arbitrary. Thus in order that (\ref{quantize}) depends only on the homology class of 
$\Sigma_A\in H_{2n-1}(Y,\R)$, the right hand side of (\ref{consistent}) must be zero for all $\lambda$. But this is 
precisely the condition (\ref{constraint}). This is perhaps not surprising: the constraint (\ref{constraint}) is 
necessary (but not sufficient) for $F_5\mid_{Y_7}$ and $*_{11} G$ to be closed. 

Provided we 
only use representatives of $(2n-1)$-cycles that are tangent to $\xi$,  and that (\ref{constraint}) also holds, 
equation (\ref{quantize})
makes sense as a topological flux quantization condition for our supersymmetric geometries. With this understanding, the left hand side 
depends only on the homology class of $\Sigma_A$, and the choice of vector field $\xi$ and 
basic K\"ahler class. 
We will see many of the above general features exemplified in more detail in sections \ref{sec:examplesT2} and \ref{sec:examples}.

\subsection{Extremal problem and the central charge}\label{sec:c}

With all of the above background now in place, we can finally summarize the extremal problem
of interest. For the case of $n=3$ we will relate this to the central charge of the dual SCFT, hence giving
a geometric dual of $c$-extremization. When $n=4$ we will show that the extremization problem allows one
to determine the two-dimensional Newton constant, $G_2$, which is related to the partition function of the dual quantum mechanics and also with the entropy of certain black hole solutions in AdS$_4$.

Much as in \cite{Martelli:2006yb}, we fix a complex cone $C(Y_{2n+1})=\R_{>0}\times Y_{2n+1}$ 
with holomorphic volume form, and holomorphic $U(1)^s$ action. A general choice of R-symmetry vector may then 
be written as in (\ref{trialR}), under which the holomorphic volume form has fixed charge $1/\cc = 2/(n-2)$. 
For a particular choice of $\xi$ and hence foliation 
$\mathcal{F}_\xi$ we may then choose a transverse K\"ahler metric with 
basic class $[J]\in H^{1,1}_{{B}}(\mathcal{F}_\xi)$. 
Finally, we should also impose flux quantization, which 
requires us to first impose the constraint (\ref{constraint}), and then (\ref{quantize}) for the cases of $n=3,4$.
For $n>4$, one could impose (\ref{quantize}) with an arbitrary constant factor multiplying $N_A$ on the right hand side.
These latter conditions will, in general, further constrain the choice of $\xi$ and $[J]$. 
By construction, a solution to the equations of motion will be a critical point 
of the action (\ref{Ssusy}), where we vary over the remaining unconstrained variables 
in $\xi$ and $[J]$. We will see how to impose all of this concretely in a class 
of examples in section \ref{sec:examplesT2}.

Geometrically we have set up a very analogous problem to volume minimization in 
Sasakian geometry \cite{Martelli:2005tp, Martelli:2006yb}, which for Sasaki-Einstein 
five-manifolds is a geometric dual of $a$-maximization in the dual four-dimensional 
SCFTs. It is thus natural to interpret the above, when $n=3$, as a geometric dual 
to $c$-extremization, which is a precise analogue for 
two-dimensional $(0,2)$ SCFTs \cite{Benini:2012cz}. 
However, for this analogy to hold our extremal function 
(\ref{Ssusy}) should play the role of a trial central charge function, 
and in particular be equal to the central charge of the solution at a critical point.

To see that this is indeed the case, we begin by recalling the general formula for the central charge
of the $d=2$ SCFT,
\bea
\csugra &=& \frac{3L}{2G_3}~,
\eea
where $G_3$ denotes the effective Newton constant in three dimensions. 
This is easily computed via dimensional reduction for the class of IIB backgrounds 
we are considering when $n=3$ ({\it e.g.} see appendix B of \cite{Benini:2015bwz}), 
and one finds
\bea\label{cSUGRAgen}
\frac{1}{G_3}&=& \frac{L^7}{G_{10}}\int_{Y_7} \ex^{-2B}\, \vol_7~,
\eea
where the ten-dimensional Newton constant is
\bea\label{ncten}
G_{10} &=& \frac{(2\pi)^7g_s^2\ell_s^8}{16\pi}~.
\eea
Evaluating the general expression (\ref{cSUGRAgen}) for our class of off-shell supersymmetric geometries, 
we can define what we will call the ``trial central charge", $\cZ$, via
\bea\label{cS}
\cZ & \equiv & \frac{3L^8}{(2\pi)^6g_s^2\ell_s^8} \Ssusy~,
\eea
where $\Ssusy$ is the supersymmetric action (\ref{Ssusy}) with $n=3$. Then for an on-shell supersymmetric solution
we get
\bea\label{cS2}
\cZ  |_\mathrm{on-shell} & = & \csugra~.
\eea
This completes our identification of a geometric version of $c$-extremization.

We can also consider the extremization problem for AdS$_2$ solutions of $D=11$ supergravity by setting $n=4$.
We define the two-dimensional Newton constant $G_2$ by
\bea\label{G2gen}
\frac{1}{G_2} &=& \frac{L^9}{G_{11}}\int_{Y_9} \ex^{-3B}\, \vol_9~,
\eea
where the eleven-dimensional Newton constant is
\bea\label{ncel}
G_{11} &=& \frac{(2\pi)^8\ell_p^9}{16\pi}~.
\eea
Evaluating (\ref{G2gen}) for our class of off-shell supersymmetric geometries, we have
\bea
\frac{1}{G_2} &=& \frac{16\pi L^9}{(2\pi)^8\ell_p^9} \int_{Y_9}\ex^B\, \eta\wedge \frac{J^4}{4!} \ = \ \frac{16\pi L^9}{(2\pi)^8\ell_p^9}\, \Ssusy~.
\eea
Thus, our extremal problem allows us to determine $G_2$. 

While AdS$_2$ holography is still being developed ({\it e.g.} 
\cite{Bena:2018bbd} and references therein), it is clear that $G_2$ encodes important information of the dual
superconformal quantum mechanics. A simple dimensional reduction of $D=11$ supergravity
action to $D=2$, leads to an action of the form $\frac{1}{16\pi G_2}\int \diff ^2 x\sqrt{-g_2}[R_2+\dots]$. 
This action does not have any gravitational dynamics and in particular does not give rise to an AdS$_2$ vacuum, 
so we should therefore include additional minimal degrees of freedom in the reduction as discussed in \cite{Castro:2014ima,Cvetic:2016eiv}, for example. In any event we note that the renormalized action $\frac{1}{16\pi G_2}\int_{M} \diff ^2 x\sqrt{-g_2}R_2+\frac{1}{8\pi G_2}\int_{\partial M} K$, where $K$ is the trace of the extrinsic curvature, evaluates on the unit radius Euclidean AdS$_2$ vacuum, {\it i.e.} the hyperbolic disc, to give $-1/4G_2$, and it is therefore natural to identify this as minus the logarithm of the partition function for a one-dimensional dual superconformal quantum mechanics. 

There is a special subclass of AdS$_2\times Y_9$ solutions where we can make a more precise statement 
and also make an interesting connection with black hole entropy computations for certain AdS$_4$ black hole solutions.
To see this we first recall that there is a consistent Kaluza-Klein reduction
of $D=11$ supergravity on an arbitrary $SE_7$ to obtain minimal gauged supergravity theory in $D=4$ \cite{Gauntlett:2007ma}. 
This latter
theory admits supersymmetric AdS$_4$ black hole solutions with black hole horizon given by AdS$_2 \times \Sigma_g$,
where $\Sigma_g$ is a Riemann surface with genus $g>1$ \cite{Caldarelli:1998hg}. After uplifting the
latter on $SE_7$ we obtain precisely a special example of the $D=11$ solutions we are considering in this paper
with eight-dimensional K\"ahler base given by $\Sigma_g\times KE_6$, where $KE_6$ is the transverse K\"ahler-Einstein metric
associated with the $SE_7$ metric. A short calculation, that we have included in appendix \ref{norm}, shows that
the Bekenstein-Hawking entropy, $S_{BH}$, of this class of black hole solutions is directly related to $G_2$ via
\begin{align}
S_{BH} \ = \ \frac{1}{4G_2}~.
\end{align}
Thus, for this class of black hole solutions our variational problem gives rise to the black hole entropy. Furthermore,
this is also the logarithm of the twisted topological index for the 
$\mathcal{N}=2$ SCFT in $d=3$, dual to AdS$_4\times SE_7$, after compactification on the Riemann surface $\Sigma_g$, with the addition
of R--symmetry magnetic flux (only) on $\Sigma_g$, the ``universal twist" \cite{Azzurli:2017kxo}. 

Finally, recall that the AdS$_2\times Y_9$ solutions with $Y_9=T^2\times Y_7$
can be dimensionally reduced and then T-dualized to obtain 
AdS$_3\times Y_7$ solutions of type IIB \cite{Gauntlett:2006ns}. In appendix~\ref{dimredtdual} we also derive the relationship
between $\csugra $ and $G_2$ for this class of solutions.


\section{A class of examples: $Y_7=T^2\times Y_5$}\label{sec:examplesT2}

In this section we use the general formalism described in section \ref{sec:general} 
to discuss a class of type IIB AdS$_3\times Y_7$ examples in which $Y_7=T^2\times Y_5$ is a product of a flat two-torus 
$T^2$ with a compact five-manifold $Y_5$. As noted earlier, after T-duality and uplifting to $D=11$ these give
AdS$_2\times T^4\times Y_5$ examples (see appendix \ref{dimredtdual}). 
We will focus on the type IIB perspective in the remainder of the paper.

We begin in section \ref{sec:geometry} 
by specializing the formulae of section \ref{sec:general} to this case. In 
particular, the main simplifying feature is that the dependence of $\Ssusy$ on the K\"ahler class 
parameters can be entirely eliminated using flux quantization, so that the resulting 
extremal function is a function only of the R-symmetry vector $\xi$. In section~\ref{sec:fixed} 
we describe a general localization formula that allows one to compute this extremal function 
quite explicitly in terms of fixed point data of $\xi$. In section~\ref{sec:obstruction}
 we show that complex cones 
$C(Y_5)\equiv \R_{>0}\times Y_5$ which 
admit a compatible K\"ahler cone metric \emph{never} have a  corresponding 
AdS$_3\times T^2\times Y_5$ solution. This implies that the complex geometry 
of $C(Y_{5})$ for AdS$_3\times T^2\times Y_5$ solutions is necessarily somewhat exotic, 
and we describe this in further detail in section \ref{sec:toric}, and in
the explicit examples in sections \ref{sec:Ypq}
and \ref{sec:Labc}. The result of section \ref{sec:obstruction} also implies 
that compactifying four-dimensional SCFTs, dual to AdS$_5\times Y_5$ Sasaki-Einstein solutions,
on $T^2$ with no geometric twist cannot flow to two-dimensional $(0,2)$ SCFTs with 
AdS$_3$ duals of the type studied in this paper, if the complex structure on the cones is preserved in the RG flow.

\subsection{General formulas}\label{sec:geometry}

Throughout this section we will assume that $Y_7=T^2\times Y_5$, where in addition $b_1(Y_5)\equiv \dim H_1(Y_5,\R)=0$ and 
 the R-symmetry 
vector $\xi$ is taken to be tangent to $Y_5$. We then write the transverse K\"ahler form as\footnote{In fact we 
only need this equation to hold in basic cohomology.}
\bea
J &=& A\, \vol_2 + \omega~,
\eea
where the volume form $\vol_2$ on $T^2$ is normalized so that $\int_{T^2}\vol_2 = 1$, and $A>0$ is a constant 
parametrizing the K\"ahler class of $T^2$. The two-form $\omega$ is a transverse K\"ahler form  on $Y_5$. 
Notice that because the $T^2$ is flat, the transverse K\"ahler metric on $Y_5$ solves \eqref{boxR}.
Notice also that $\rho=\diff P$ is then similarly a transverse Ricci form on $Y_5$, again since the torus is flat. 
For this class 
the supersymmetric action (\ref{Ssusy}) is easily computed, giving
\bea\label{Ssusy5}
\Ssusy &=& A \int_{Y_5}\eta\wedge \rho \wedge \omega~.
\eea
Cancelling an overall factor of $A\neq 0$, the constraint (\ref{constraint}) similarly reads
\bea\label{constraint5}
\int_{Y_5}\eta \wedge \rho \wedge \rho &=& 0~.
\eea

We next turn to flux quantization. Recall that we assume $b_1(Y_5)=0$. The K\"unneth formula then implies that the 
five-cycles on $Y_7$ are spanned by a copy of $Y_5$ at a fixed point on $T^2$, and $\Sigma_I=T^2 \times \sigma_I$, where 
$\sigma_I\subset Y_5$ form a basis of three-cycles in $Y_5$, $I=1,\ldots,b_3(Y_5)\equiv \dim H_3(Y_5,\R)$. 
Noting that $\xi$ is tangent to $Y_5$, (\ref{quantize}) reads
\bea\label{omegaN}
\int_{Y_5} \eta\wedge \rho \wedge \omega &=& \frac{2(2\pi\ell_s)^4g_s}{L^4}N~,
\eea
where we have denoted the flux number for this distinguished five-cycle by $N \in \mathbb{N}$.
Substituting into the off-shell supersymmetric action (\ref{Ssusy5}) immediately gives
\bea\label{SsusyN}
\Ssusy &=& A\frac{2(2\pi\ell_s)^4g_s}{L^4}N~.
\eea

We next claim that there do not exist solutions with $b_3(Y_5)=0$, for example ruling out $Y_5=S^5$ topology. 
Recall that $\xi$ leads to a foliation $\mathcal{F}_\xi$ of $Y_5$, and there is an associated long exact Gysin sequence. The relevant 
part of this for our purposes reads
\bea
0  \ \cong \ H^3(Y_5,\R) \longrightarrow H^2_B(\mathcal{F}_\xi) \longrightarrow H^4_B(\mathcal{F}_\xi) \longrightarrow H^4(Y_5,\R) \ \cong \ 0~.
\eea
Here we have used $b_1(Y_5)=0=b_3(Y_5)$. 
This sequence implies $H^2_B(\mathcal{F}_\xi)\cong H^4_B(\mathcal{F}_\xi)\cong \R$, the latter being generated by the transverse volume form. On the other hand 
the foliation is transversely K\"ahler, so the transverse K\"ahler class must generate the former group. 
It follows that $[\rho]=\lambda [\omega]\in H^2_B(\mathcal{F}_\xi)$ for some constant $\lambda\in \R$. 
But the constraint (\ref{constraint5}) then implies
\bea
0 &=& \int_{Y_5} \eta\wedge \rho \wedge \rho \ = \ \lambda^2 \int_{Y_5}\eta\wedge \omega\wedge \omega \ = \ 2\lambda^2\int_{Y_5}\eta \wedge \vol_4~,
\eea
which implies $\lambda=0$, a contradiction. Indeed, note that the action (\ref{Ssusy5}) is then zero. 
Notice that when $b_3(Y_5)=0$ there is by definition no baryonic $U(1)$ symmetry in the 
dual field theory. 

We next look at the case $b_3(Y_5)=1$, so there is a single three-cycle generated by $\sigma\subset Y_5$. 
Under our above topological assumptions one can similarly show that the Gysin long exact sequence implies 
\bea
H^2(Y_5,\R) &\cong & H^2_B(\mathcal{F}_\xi)/[\, \rho\, ]~.\label{Gysin}
\eea
This in turn implies (\ref{topcondition}). 
It follows that $\sigma$ may be taken to be tangent to the R-symmetry vector. The quantization condition (\ref{quantize}) then
reads\footnote{\label{foot9}Much of the above analysis also applies, {\it mutatis mutandis}, for $D=11$ solutions with $Y_9=T^2\times Y_7$, including the conclusion that there do not exist solutions with $b_5(Y_7)=0$. However, the analogue of the integral in \eqref{aM} will involve $\eta\wedge\rho\wedge\omega$, where $\omega$ is the transverse K\"ahler form on $Y_7$ and hence the flux quantization conditions still depend on the transverse K\"ahler class when $Y_9=T^2\times Y_7$.}
\bea
A\int_\sigma \eta\wedge \rho &=& \frac{2(2\pi\ell_s)^4g_s}{L^4}M~,\label{aM}
\eea
where we have denoted the flux quantum number of the five-cycle $T^2\times \sigma$ as $M\in\mathbb{Z}$. 
Putting everything together, we get the following very simple expression for what we will call the ``trial''  central charge
\bea
 \cZ &\equiv & \frac{12(2\pi)^2 MN}{\int_\sigma \eta\wedge \rho}~.\label{cSUGRAb31}
\eea
The numerator is of course quantized. The flux quantization conditions have effectively allowed us 
to eliminate the dependence on the K\"ahler class in terms of the integers $N$ and $M$, 
and the only 
 dependence on the R-symmetry vector $\xi$ is now purely in the denominator. 
We should thus now extremize (\ref{cSUGRAb31}) as a function of $\xi$, subject to the constraint 
(\ref{constraint5}).

Finally, for general $b_3(Y_5)\geq 1$ the formula (\ref{cSUGRAb31})  of course still holds, where 
we pick one of the generating three-cycles to be $\sigma\equiv \sigma_1$, with corresponding flux quantum number 
$M\equiv M_1$. However, we must in addition impose flux quantization through each $\sigma_I$, $I=2,\ldots,b_3(Y_5)$, which is equivalent to imposing
\bea\label{ratios}
\frac{\int_{\sigma_I}\eta\wedge \rho}{\int_{\sigma}\eta \wedge \rho} &=& \frac{M_I}{M}~, \qquad I\ = \ 2,\ldots,b_3(Y_5)~.
\eea
The flux quantum numbers $\{M=M_1,M_2,\ldots,M_{b_3(Y_5)}\}$ are part of the fixed global, topological data.

\subsection{Fixed point theorem}\label{sec:fixed}

In practice we would like to obtain more explicit expressions for quantities such as 
(\ref{constraint5}) and  (\ref{cSUGRAb31}). To do this we may use similar 
techniques to those in \cite{Martelli:2006yb}. For generality we again return to general complex dimension 
$n$, with $n=3$ being the case relevant for type IIB supergravity solutions\footnote{It is worth noting that for the $D=11$ AdS$_2\times Y_9$ solutions with $Y_9=T^2\times Y_7$, or equivalently, AdS$_3\times Y_7$ type IIB solutions,
one needs to calculate different integrals, involving the transverse 
K\"ahler class of $Y_7$, as noted in footnote \ref{foot9}.}. Thus associated to $Y_{2n-1}$ we have a complex cone $C(Y_{2n-1})=\R_{>0}\times Y_{2n-1}$, with coordinate $r>0$ 
on the first factor. On $C(Y_{2n-1})$ we then introduce the two-form
\bea\label{gamma}
\gamma & \equiv & \frac{1}{2}\diff (r^2\eta) \ = \ r\diff r\wedge \eta + \frac{1}{2}r^2\diff \eta~.
\eea
It is straightforward to show that
\bea\label{trick}
\int_{C(Y_{2n-1})} \ex^{-\frac{1}{2}r^2 + \gamma} &=& \int_{Y_{2n-1}} \eta \wedge (\diff \eta)^{n-1}~. 
\eea
This equality follows simply by explicitly performing the integral over $r\in (0,\infty)$ on the cone.
Since 
\bea
\diff \left(-\tfrac{1}{2}r^2\right) & = & \xi\lrcorner \gamma~,
\eea
it follows that $-\frac{r^2}{2} + \gamma$ is an equivariantly closed form under the 
derivative $\diff - \xi\lrcorner$. 
The left hand side of (\ref{trick}) may hence be computed using the Berline-Vergne
localization formula. However, as in the similar application of the Duistermaat-Heckman
formula to Sasakian geometry in \cite{Martelli:2006yb}, where $\gamma$ is a symplectic form, the fixed point set 
of $\xi$ is formally the origin $r=0$. This is generically a singular point if we add it 
to compactify $C(Y_{2n-1})$ around $r=0$, and the Berline-Vergne formula cannot be applied.

We may obtain a meaningful formula by instead \emph{resolving} the 
 singularity at the origin, 
again as in \cite{Martelli:2006yb}. There is no unique way to do this, and the resulting formulas 
will take  different forms for different resolutions. By a (partial) resolution here we mean a manifold (or respectively orbifold)
$\hat{C}(Y_{2n-1})$ together with a map $\pi: \hat{C}(Y_{2n-1})\rightarrow C(Y_{2n-1})\cup \{r=0\}$. This map should be 
equivariant under the holomorphic $U(1)^s$ action, and be a diffeomorphism 
when restricted to $C(Y_{2n-1})$. We refer to $\pi^{-1}(\{r=0\})$ as the \emph{exceptional set}. 
Since by definition $\xi$ is nowhere zero on $C(Y_{2n-1})$, its fixed point set is a subset of the exceptional 
set. 
There is a canonical way to construct such a partial resolution: simply pick 
a quasi-regular R-symmetry vector field $\xi_0$, and take $\hat{C}(Y_{2n-1})$ to be 
the total space of the orbifold line bundle $K^{\mm/I_{V_0}}$, described at the end of section 
\ref{sec:cone}. In this case the exceptional set is a copy of the K\"ahler orbifold $V_0$, which maps to 
$\{r=0\}$ under the map $\pi$. The fixed point set of $\xi_0$ is precisely $V_0$, but 
for a more general R-symmetry vector (\ref{trialR}) there will be an induced action 
of $U(1)^{s-1}$ on the exceptional set, where for generic $\xi$ its fixed points will coincide with the fixed points of $U(1)^{s-1}$.
On this particular partial resolution $\hat{C}(Y_{2n-1})$ we may identify 
$r$ with a radial distance function on the complex line fibre. 
Notice that although $\eta$ is not 
defined on the exceptional set, where $\xi$ has fixed points, 
nevertheless the two-form $\gamma$ defined in (\ref{gamma}) is well-defined everywhere 
when pulled back to $\hat{C}(Y_{2n-1})$ -- it is simply zero on the exceptional set.

With this notation in hand, we may then apply the Berline-Vergne fixed point theorem 
on $\hat{C}(Y_{2n-1})$, which gives
\bea
\int_{Y_{2n-1}} \eta \wedge (\diff \eta)^{n-1} \ = \  (2\pi)^n\sum_{\{\sF\}} \prod_{\lambda=1}^R \frac{1}{(\vec{b},\u_\lambda)^{n_\lambda}} \int_\sF \frac{1}{d_\sF} \prod_{\lambda=1}^R \left[\sum_{a\geq 0}\frac{c_a(\mathcal{E}_\lambda)}{(\vec{b},\u_\lambda)^a}\right]^{-1}.\label{localize}
\eea
The notation here is exactly the same as in \cite{Martelli:2006yb}. The sum is over connected components
$\sF$ of the fixed point 
set of $\xi$ on $\hat{C}(Y_{2n-1})$. As described above, each such $\sF$ is a subset of the exceptional set $\pi^{-1}(\{r=0\})$. 
For each connected component $\sF$ of fixed points 
 the $\u_\lambda\in\mathbb{Z}^s\subset \mathtt{t}_s^*$ denote weights of the resulting linear action of 
$\xi$ on the normal bundle, with multiplicities $n_\lambda\in\mathbb{N}$, so that the linear action by $\xi$ on a 
given weight space $\mathcal{E}_\lambda$ is $(\vec{b},\u_\lambda)$. Here 
we use the basis for the Lie algebra $\mathtt{t}_s$ of $U(1)^s$ 
in (\ref{trialR}), so that  $\vec{b}=(b_1,\ldots,b_s)$ parametrizes the choice of R-symmetry vector. 
Finally $c_a(\mathcal{E}_\lambda)$ denote Chern classes of the corresponding weight space bundles, 
and the positive integer 
$d_\sF$ is the \emph{order} of $\sF$ as an orbifold, and is required only for resolutions with orbifold singularities. 

Although the general formula (\ref{localize}) is a little cumbersome, the point is that the right hand 
side is manifestly only a function of the trial R-symmetry vector $\vec{b}$, together with certain global topological data -- 
namely Chern numbers and weights of the $U(1)^s$ action. If $\hat{C}(Y_{2n-1})$ is a smooth manifold with only 
isolated fixed points of $U(1)^n$ (a maximal torus action), the right hand side significantly simplifies to
\bea\label{fpf}
\int_{Y_{2n-1}} \eta \wedge (\diff \eta)^{n-1} &=& (2\pi)^n\sum_{\{\mathrm{fixed\, points}\}}\prod_{\lambda=1}^n \frac{1}{(\vec{b},\u_\lambda)}~,
\eea 
where the $n$, possibly indistinct, weights are $\u_\lambda$, $\lambda=1,\ldots,n$. 
Recalling that $\diff\eta = \cc \rho = \tfrac{1}{2}\rho$ when $n=3$, the 
formula (\ref{localize}) similarly implies that  the constraint (\ref{constraint5}) and central charge (\ref{cSUGRAb31}) depend only 
on the trial R-symmetry vector $\vec{b}$ and global, topological data.
In particular we may apply a  similar localization formula (\ref{localize}), simply replacing $Y_{2n-1}$ by $\sigma$. Here we 
take $\sigma\subset Y_5$ to be a three-submanifold, invariant under $U(1)^s$. The resolution $\hat{C}(Y_{5})$ of 
$C(Y_{5})\cup\{r=0\}$ will induce a resolution of the cone over $\sigma$, although again the point is that 
we may use \emph{any} choice of resolution. 

We shall make use of the general formulae in this section in the examples of section~\ref{sec:examples}.

\subsection{An obstruction for K\"ahler cones}\label{sec:obstruction}

Let $Y_5$ be a five-manifold with complex cone $C(Y_{5})=\R_{>0}\times Y_5$ that is of 
Calabi-Yau type. By this we mean that the complex manifold $C(Y_{5})\cong \R_{>0}\times Y_5$ admits a K\"ahler cone metric that is compatible with the given complex structure, 
and has a global holomorphic volume form of positive charge under the Reeb vector. In this section we show that 
there is no supersymmetric AdS$_3\times T^2 \times Y_5$ solution with the given complex structure on 
$C(Y_{5})$. In fact as we shall see, the problem is quite simple:  an R-symmetry vector $\xi$ satisfying 
the constraint (\ref{constraint5}) necessarily lies outside the \emph{Reeb cone} (defined below)
and in that 
case the putative radial vector $r\partial_r \equiv -\mathcal{I}(\xi)$ has no compatible 
radial coordinate $r>0$. Thus, in a sense the complex geometry is not compatible with the radial 
slicing into $C(Y_{5})=\R_{>0}\times Y_5$.
For simplicity we shall prove this for toric K\"ahler cones, using some of the formalism of \cite{Martelli:2005tp}, 
although the proof can be generalized. 

Thus let $C(Y_{5})$ be a toric complex cone of Calabi-Yau type. Fix a choice of any compatible K\"ahler cone metric. Following reference \cite{Martelli:2005tp} we may then introduce 
symplectic-toric coordinates $(y_1,y_2,y_3;\varphi_1,\varphi_2,\varphi_3)$, where the Killing vectors
$\partial_{\varphi_i}$, $i=1,2,3$, generate the effectively acting $U(1)^3$ action. The 
coordinates $y_i$ arise as moment maps, and lie inside a convex polyhedral cone 
$\vec{y}\in \mathcal{C}\subset \R^3\cong \mathtt{t}^*_3$, where $\mathtt{t}_3$ is the Lie algebra of the torus $U(1)^3$.
The K\"ahler cone metric may be written
\bea
\diff s^2_{\mathrm{Kahler}} &=& \sum_{i,j=1}^3 G_{ij}\diff y_i\diff y_j + G^{ij}\diff\varphi_i\diff\varphi_j~,
\eea
where $G_{ij}=G_{ij}(\vec{y})$ is homogeneous degree $-1$, and is positive definite and smooth in the interior of $\mathcal{C}$, 
with a certain pole behaviour on the bounding facets of $\mathcal{C}$, required in order that the metric 
compactifies smoothly there. Here $G^{ij}$ is simply the inverse matrix to $G_{ij}$, which correspondingly 
has reduced rank on the boundary components of $\mathcal{C}$. 
This K\"ahler cone metric will have an associated Reeb vector field 
\bea
\mathrm{Reeb} &=& \sum_{i,j=1}^3 2G_{ij}y_j\partial_{\varphi_i}~.
\eea
However, this will \emph{not} be a putative R-symmetry vector in the application to AdS$_3$ solutions, as we describe 
below. Instead we wish to use the above coordinates simply to describe the complex geometry of $C(Y_{5})$, rather than 
the K\"ahler cone geometry that is also present in the above description.

Let
\bea\label{Rvector}
\xi &=& \sum_{i=1}^3 b_i \partial_{\varphi_i}
\eea
be a putative R-symmetry vector for a supersymmetric AdS$_3\times T^2\times Y_5$ solution. 
Recall that the 
\emph{Reeb cone} $\mathcal{C}^*\subset \R^3\cong \mathtt{t}_3$ is
defined as $\mathcal{C}^*=\{\vec{b}\in \R^3\mid (\vec{b},\u_\alpha)>0\}$, where
\bea\label{Cu}
\mathcal{C}=\left\{\sum_{\alpha}t_\alpha \u_\alpha\in \R^3\mid t_\alpha\geq 0\right\}~.
\eea
The $\u_\alpha\in\Z^3$ are the outward-pointing generating edges of the moment map polyhedral cone $\mathcal{C}$, 
and for the Reeb vector of a K\"ahler cone metric one necessarily has $\xi \in \mathcal{C}^*$ \cite{Martelli:2005tp}. 
Moreover, for such a vector 
\bea\label{Sasvol}
\int_{Y_5}\eta\wedge \rho \wedge \rho &=& 32\,  \mathrm{Vol}(\xi) \ > \ 0~,
\eea
where $\mathrm{Vol}(\xi)$ denotes the Riemannian volume of the corresponding Sasakian metric. 
Thus for $\xi\in\mathcal{C}^*$ the constraint (\ref{constraint5}) cannot hold, since this requires the 
left hand side of (\ref{Sasvol}) to be zero. Thus $\xi\notin\mathcal{C}^*$, meaning there 
is at least one edge vector $u_\alpha$ with
\bea\label{alphaout}
(\vec{b},\u_\alpha) &< & 0~.
\eea
Geometrically, the edge vector $\u_\alpha$ corresponds to a one-dimensional torus-invariant complex submanifold 
$V_\alpha \cong \C^*=U(1)\times \R_{>0}$ 
of $C(Y_{5})$. Here the $U(1)$ is the single non-vanishing circle over that edge in $\partial\mathcal{C}$. 
Being torus-invariant, the R-symmetry vector (\ref{Rvector}) is tangent to $V_\alpha$, and so too is it's complex partner
\bea
-\mathcal{I}(\xi) &=& \sum_{i,j=1}^3 G^{ij}b_j\frac{\partial}{\partial y_i} \ = \ r\partial_r~.
\eea
In the second equality we have used the formula for the complex structure in symplectic-toric coordinates (equation (2.16) of \cite{Martelli:2005tp}), 
while the last equality follows from the relationship between the radial vector and Reeb vector 
described in section \ref{sec:cone}. Note that when restricted to an edge vector $\vec{y}\in \{t\, \u_\alpha\mid t \geq 0\}\subset \partial\mathcal{C}$, 
the matrix $G^{ij}(\vec{y})$ has rank 1, corresponding to the single $U(1)$ that is non-vanishing along the pre-image of that edge 
under the moment map.
On the other hand, the outward-pointing 
directional derivative along the edge corresponding to $\u_\alpha$ is by definition
\bea
\nu_\alpha \ \equiv \ \sum_{i=1}^3 u_\alpha^i\frac{\partial}{\partial y_i}~.
\eea 
We thus see that
\bea
\sum_{j=1}^3 G^{ij}b_j & \propto & u_\alpha^i
\eea
holds along the edge generated by the vector $\u_\alpha$. We may determine the proportionality factor by dotting this with $\vec{b}$. Note immediately that 
the right hand side is negative due to (\ref{alphaout}), while the left hand side is $\sum_{i,j=1}^3 G^{ij}b_ib_j=|\xi|^2>0$, the square length of the R-symmetry vector. 
Thus along the edge vector generated by $\u_\alpha$, namely $\vec{y}\in \{t\,  \u_\alpha\mid t \geq 0\}\subset \partial\mathcal{C}$, 
we have shown that
\bea\label{Eulerray}
r \partial_r &=& \sum_{i,j=1}^3 G^{ij}b_j\frac{\partial}{\partial y_i} \ = \ \frac{|\xi|^2}{(\vec{b},\u_\alpha)}\nu_\alpha~.
\eea
Thus when the R-symmetry vector $\xi$ lies outside the Reeb cone, so that (\ref{alphaout}) holds for some $\alpha$, 
there is a corresponding submanifold $V_\alpha\subset C(Y_{5})$ along which $r\partial_r$ points \emph{towards} the origin, rather than 
away from it. This is an immediate problem given the definition of the radial coordinate $r$. By definition we have
\bea\label{homogen}
r \partial_r r^2 &=& 2r^2 >0~,
\eea
away from the origin $r=0$. On the other hand, combining this with (\ref{Eulerray}) says that $r^2$ is monotonic \emph{decreasing} 
as one moves out along the edge, {\it i.e.} as $t$ increases from zero in $\{t\, \u_\alpha \mid t\geq 0\}$. This is a contradiction, 
since $r^2=0$ at the tip of the cone $t=0$, and should clearly by non-negative.

This concludes our proof, but since it is rather general (and abstract), it is perhaps helpful to give a simple example where 
the details can be seen more explicitly. Thus consider $C(Y_{5})=\C^3\setminus \{0\}$. In this case we may introduce 
polar coordinates $(r_i,\varphi_i)$, $i=1,\ldots,3$, for each copy of $\C$ in $\C^3=\oplus_{i=1}^3 \C$. 
The moment map coordinates are $y_i=\frac{1}{2}r_i^2\geq 0$, so that the polyhedral cone is
$\mathcal{C}=(\R_{\geq 0})^3\cong \mathcal{C}^*$. The matrix $G_{ij}$ and its inverse $G^{ij}$ for the flat K\"ahler metric on $\C^3$ are
\bea
G_{ij} &=& \mathrm{diag}\left(\frac{1}{r_1^2},\frac{1}{r_2^2},\frac{1}{r_3^2}\right)~, \qquad G^{ij} \ = \ \mathrm{diag}(r_1^2,r_2^2, r_3^2)~.
\eea
The $3$ bounding facets of $\mathcal{C}$ are at $\{r_i=0\}$, $i=1,2,3$, where notice that $G^{ij}$ has a zero. The 
generating edge vectors are $\u_1=(1,0,0)$, $\u_2=(0,1,0)$, $\u_3=(0,0,1)$, where the edge corresponding to 
$\u_1$ is $\{r_2=0,r_3=0\}$, {\it etc}. Suppose the R-symmetry vector lies outside the Reeb cone, 
which means that $b_i<0$ for at least one $i=1,2,3$. Without loss of generality, let us suppose that $b_1<0$. 
Then along $V_1\equiv C(Y_{5})\cap \{r_2=r_3=0\}\cong \C^*$ the radial vector is
\bea
r\partial_r &=& \sum_{i=1}^3 b_i r_i\partial_{r_i}\ = \ b_1 r_1\partial_{r_1}~.
\eea
But since $b_1<0$, this says that $r^2$ is a monotonic decreasing function of the radius $r_1$ along $V_1$.
For example, the expression $r^2=\sum_{i=1}^3 r_i^{2/b_i}$ satisfies the homogeneity equation (\ref{homogen}), 
and indeed notice that for $b_1<0$ it is a monotonic decreasing function of $r_1$ along $\{r_2=0, r_3=0\}$, as we have 
argued in general. However, $r_1=0$ maps to $r=\infty$ for this choice of ``radial'' coordinate! 

\subsection{Toric formulas and non-convex cones}\label{sec:toric}

In the previous subsection we have ruled out toric complex cones $C(Y_5)$ of Calabi-Yau type as giving rise to AdS$_3\times T^2\times Y_5$ solutions. 
However, one can nevertheless write down formulas for the physical quantities of interest in section \ref{sec:geometry} in this case, that we will use in the next section. Furthermore, they will lead us to some formulae, which we conjecture to hold
for ``non-convex" toric geometries (defined below), that are associated with some explicitly known AdS$_3\times T^2\times Y_5$ solutions.

We begin with the toric complex cones $C(Y_5)$ with a compatible K\"ahler cone metric, as in section \ref{sec:obstruction}.
Denoting $d\geq 3$ as the number of facets of the associated polyhedral cone $\mathcal{C}$, we may rewrite the 
presentation of $\mathcal{C}$ in (\ref{Cu}) via a dual description as
\bea\label{Cv}
\mathcal{C} &=& \{\vec{y}\in \R^3 \mid (\vec{y},\v_a) \geq 0~, \ a \ = \ 1,\ldots, d\}~.
\eea
Here $\v_a\in \Z^3$ are the \emph{inward}-pointing normal
vectors to the $d$ facets of the cone. Each facet 
of $\mathcal{C}$, namely $\{(\vec{y},\v_a)=0\}$, corresponds to a complex codimension one submanifold, which is a 
cone over a three-manifold $S_a\subset Y_5$. We then have the formula~\cite{Martelli:2005tp}
\bea\label{toric3}
\int_{S_a} \eta\wedge \rho &=&  2(2\pi)^2\frac{(\v_{a-1},\v_a,\v_{a+1})}{(\vec{b},\v_{a-1},\v_a)(\vec{b},\v_a,\v_{a+1})}~.
\eea
Similarly, it is a standard result of toric geometry ({\it e.g.} see eq. (17) of \cite{Abreu:2010zz}) 
that the transverse basic first Chern class
\bea\label{c1toric}
c_1^B &=& \frac{[\rho]}{2\pi} \ = \ \sum_{a=1}^d [S_a]_{\mbox{\tiny{Poincar\'e dual}}} \in H^2_{B}(\mathcal{F}_\xi)~.
\eea
Using (\ref{c1toric}) and (\ref{toric3}) we may then derive the formula
\bea\label{toricconstraint}
\int_{Y_5}\eta\wedge \rho\wedge\rho &=& \frac{4 (2\pi)^3}{b_1}\sum_{a=1}^d \frac{(\v_{a-1},\v_a,\v_{a+1})}{(\vec{b},\v_{a-1},\v_a)(\vec{b},\v_a,\v_{a+1})}~,
\eea
where recall that $b_1=2$ is fixed by the charge of the holomorphic volume form. Note that equation (\ref{toricconstraint}) was derived 
differently in \cite{Martelli:2005tp}. 

Strictly speaking,  (\ref{toric3}) and  (\ref{toricconstraint}) were proven in \cite{Martelli:2005tp} only for $\vec{b}$ lying in the Reeb cone,
$\vec{b}\in\mathcal{C}^*$, for which there is a compatible K\"ahler cone metric. However, toric K\"ahler cones in dimension $n=3$
can always be fully resolved, for which one can also apply the fixed point formula (\ref{fpf}). On the other hand, this latter 
formula uses the general Berline-Vergne theorem, which does not require any K\"ahler cone structure, and in 
particular applies to any R-symmetry vector $\vec{b}$. Notice here that the radial coordinate $r$ in section \ref{sec:fixed} can 
be \emph{any} choice of radial coordinate on the toric K\"ahler cone $C(Y_5)$, and does not have to be paired with the R-symmetry vector via a complex structure.
Since the fixed point formulas and (\ref{toric3}), (\ref{toricconstraint}) necessarily agree for $\vec{b}\in\mathcal{C}^*$, 
being rational functions of $\vec{b}$ it follows that they agree for all $\vec{b}$.  We shall make use 
of (\ref{toric3}) and (\ref{toricconstraint}) in the examples in section \ref{sec:examples}.

In  the next section, associated with AdS$_3\times T^2\times Y_5$ solutions, we will meet examples of complex cones $C(Y_5)=\R_{>0}\times Y_5$ that have a holomorphic volume form and a holomorphic 
$U(1)^3$ action, but do \emph{not} admit any compatible K\"ahler cone metric, consistent with the obstruction theorem in the last subsection.
As for toric K\"ahler cones, one can still define 
vectors $\v_a\in\Z^3$ for such geometries: the index $a$ labels the components of torus-invariant complex codimension one submanifolds, with the vector $\v_a$ 
specifying the $U(1)\subset U(1)^3$ that fixes a given component. The vectors are normalized to be primitive, so that they define an effective 
$U(1)$ action on the normal space to the fixed point set. The signs are fixed so that minus
the complex structure pairs each vector with a radial vector that points inwards from the fixed point set  (rather than outwards). However, for the examples that we shall discuss 
the set of these $\v_a$'s  no longer define a convex polyhedral cone $\mathcal{C}$, as in (\ref{Cv}). We 
thus refer to them as ``non-convex'' toric cones. 
Notice that both sides of the equations  (\ref{toric3}) and (\ref{toricconstraint})  still make sense for this 
class of geometries. In fact we conjecture that (\ref{toric3}) and (\ref{c1toric}) continue to hold for non-convex/non-K\"ahler cones, 
although we currently have no  general proof of this.\footnote{Actually one can prove  (\ref{toric3}) by directly computing both sides, as sketched in this footnote, although we believe there should be 
a better approach. Note first, for example by choosing a quasi-regular R-symmetry vector,  that the $S_a$ are total spaces of orbifold circle bundles over a toric orbifold Riemann surface. The latter is necessarily a weighted projective space. One can then 
evaluate the left hand side using the localization formula in section \ref{sec:fixed}, partially resolving the cone $C(S_a)=\R_>0\times S_a$ to the total space of the associated 
complex line orbibundle over the weighted projective space. This gives a completely explicit formula that can be compared to the right hand side of (\ref{toric3}), to see that they agree. Notice that when the degree of the line bundle is negative these are toric and K\"ahler, for which the methods of \cite{Martelli:2005tp}, \cite{Martelli:2006yb}
may instead be used to deduce (\ref{toric3}), but for positive degree we only have the approach sketched here. } We shall instead find by explicit computation in examples 
that (\ref{toric3}) and (\ref{toricconstraint}) do always hold, where recall that (\ref{toricconstraint}) follows from  (\ref{toric3}), (\ref{c1toric}).

Finally, assuming (\ref{c1toric}) holds, we may derive another interesting formula, that is 
also valid for non-convex toric cones.
Similarly to (\ref{toricconstraint}) we may immediately deduce
\bea
\int_{Y_5}\eta \wedge \rho \wedge \omega &=& 2\pi\sum_{a=1}^d \int_{S_a} \eta\wedge \omega~,
\eea
where recall that $\omega$ is the transverse K\"ahler form on $Y_5$. On the other hand, combining this with
(\ref{omegaN}) then gives
\bea\label{2R}
2N &=& \frac{2\pi L^4}{(2\pi\ell_s)^4g_s}\sum_{a=1}^d \int_{S_a} \eta\wedge \omega~.
\eea
One can calculate R-charges of baryonic operators in the dual SCFT that are associated with D3-branes wrapping 
supersymmetric three-cycles of $Y_5\subset T^2\times Y_5$. As discussed for the explicit examples in 
\cite{Couzens:2017nnr}, that we re-examine in section \ref{sec:Ypq}, one can  associate 
such a baryonic operator to each $S_a\subset Y_5$, $a=1,\ldots,d$. 
 Moreover, we can recast the expression for
the R-charges of these operators as
\bea\label{rchgegenexpgeneral}
R[S_a] &= & \frac{L^4}{(2\pi)^3\ell_s^4g_s}\int_{S_a}\me^{-B} \vol(S_a)
\ = \ \frac{L^4}{(2\pi)^3\ell_s^4g_s}\int_{S_a}\eta\wedge \omega~,
\eea
where $\vol(S_a)$ is the volume form with respect to the metric on $Y_5$ pulled back to the three-cycle, and similarly the integrand in the second expression is understood to be pulled back to the cycle. 
Combining with (\ref{2R}) then gives
\bea\label{NR}
2N &=& \sum_{a=1}^d R[S_a]~.
\eea
We shall see via explicit computation that this indeed holds in all our examples.



\section{Examples}\label{sec:examples}

In this section we present and discuss various examples of $Y_5$ geometries that we discussed in
section \ref{sec:examplesT2}. 
Each example has quite different features, both geometrically and physically. In all of the examples the complex cones $C(Y_5)$ are toric cones, which, as we discussed in section \ref{sec:toric}, are necessarily non-convex toric for
the AdS$_3\times T^2\times Y_5$ solutions.

\subsection{$Y_5=\Ypq$}\label{sec:Ypq}

In this section we apply the above formalism to a family of examples where there is also an explicitly known
supergravity solution, first constructed in \cite{Donos:2008ug}. A proposal for the dual field theory 
was made in \cite{Benini:2015bwz}, and also subsequently discussed in \cite{Couzens:2017nnr}. Here we first extend the discussion of the geometry of these solutions and demonstrate that the proposal of \cite{Benini:2015bwz} is not correct.

From  section 4 of \cite{Donos:2008ug}, the metric on the internal space $Y_7$ is\footnote{Note that we follow the notation of \cite{Donos:2008ug}, but we have relabelled the coordinate $\gamma=z^{there}$ and also $\p=q_{\mathrm{there}}$, $\q=p_{\mathrm{there}}+q_{\mathrm{there}}$.}
\bea\label{Ypqmetric}
\diff s^2_7 &=&\frac{\diff y^2}{4\beta^2y^2 U(y)}+\frac{U(y)}{4(\beta^2-1+2y)}(\diff\psi-\cos\theta \diff\phi)^2+\frac{1}{4\beta^2}(\diff\theta^2+\sin^2\theta \diff \phi^2)\nn\\
&&+\frac{\beta^2-1+2y}{4\beta^2}D\zz^2+\frac{y}{\beta^2}\diff s^2_{T^2}~,
\eea
with $D\zz\equiv  \diff \zz- g(y)(\diff\psi-\cos\theta \diff\phi)$, and where 
\bea
U(y)&=& 1-\frac{1}{\beta^2}(1-y)^2~, \qquad  g(y) \ = \ \frac{y}{\beta^2-1+2y}~.
\eea
The parameter $\beta=p/q$, where $p,q$ are relatively prime integers satisfying
\begin{align}
q>p>0~. 
\end{align}
The ranges of coordinates are 
 $0\le\theta\le\pi/2$, $0\le\phi<2\pi$, $0\le\psi<2\pi$,
$1-\beta\le y\le 1+\beta$, $0\le \zz< 2\pi l$, with $l=2q/(q^2-p^2)$. In these coordinates, the R-symmetry vector is 
\bea\label{Ypqxi}
\xi & =& 2(\partial_\psi+\partial_\zz)~.
\eea

Topologically $Y_7=T^2\times Y_5$, where $Y_5=\Ypq\cong S^2\times S^3$. The local metric on $\Ypq$, and how it extends globally 
to a smooth metric on $S^2\times S^3$, is very analogous to the construction of the $Y^{p,q}$ Sasaki-Einstein 
manifolds \cite{Gauntlett:2004yd}. 
In particular the first line of (\ref{Ypqmetric}) is a smooth metric on a base four-manifold, realized geometrically as an $S^2$ bundle over $S^2$. 
Here the $y$ and $\psi$ coordinates are polar and azimuthal coordinates on the fibre $S^2$, respectively. 
There is then a circle bundle fibred over this base, with circle coordinate $\zz/l$, with  the integers $\p$ and $\q$
being Chern numbers describing the twisting. One may then equivalently view $\Ypq$ as the total space 
of a Lens space $S^3/\Z_{\p}$ fibred over $S^2$, where the integer $\q$ determines the twisting. 

The $\Ypq$ metrics have three commuting Killing vectors, namely $\partial_\phi$, $\partial_\psi$, $\partial_\zz$. Appropriately 
normalized, these generate a $U(1)^3$ isometry, and following the discussion in \cite{Martelli:2004wu} one can check that
\bea\label{Ypqbasis}
\partial_{\varphi_1} \, \equiv \, \partial_\psi + \frac{q-p}{2}l\partial_\zz~, \quad \partial_{\varphi_2} \, \equiv \,  -\partial_\phi - \frac{q-p}{2}l\partial_\zz~, \quad 
\partial_{\varphi_3} \, \equiv \,  \partial_\phi +\big(\frac{q-p}{2}+1\big)l\partial_\zz~,
\eea
generate an effective action of this torus. Thus each $\varphi_i$, $i=1,2,3$, has canonical period $2\pi$. Of course, this basis is unique 
only up to the action of $SL(3,\Z)$. We have chosen the basis in (\ref{Ypqbasis}) analogously to that in \cite{Martelli:2004wu} for 
the Sasaki-Einstein $Y^{p,q}$ metrics. The complex cone $C(\Ypq)$ is then {toric}, which means there 
is a holomorphic action of $(\C^*)^3$, with $U(1)^3\subset (\C^*)^3$ being the above isometry, and where the complex 
structure pairs each Killing vector with a corresponding holomorphic vector field. The holomorphic volume 
form $\Psi$ on the cone, defined in \eqref{Psidef},
has charge 1 under $\partial_\psi$, but is not charged under $\partial_\phi$ or $\partial_\zz$. 
The basis (\ref{Ypqbasis}) is thus the same as in (\ref{trialR}), with $\partial_{\varphi_2}$, $\partial_{\varphi_3}$ generating
flavour isometries.

The torus action generated by (\ref{Ypqbasis}) has fixed points, and it is straightforward to determine the location of the complex codimension one fixed point sets, 
which have maximal dimension.  There are four, each fixed by a particular $U(1)\subset U(1)^3$ determined by the vectors
\bea\label{Ypqv}
\v_1 \,  =\,  (1,0,0)~, \quad \v_2 \, = \, (1,1,0)~, \quad \v_3 \, = \, (1,p,p)~, \quad \v_4 \, = \, (1,p-q-1,p-q)~.
\eea
Using (\ref{Ypqbasis}) and the metric (\ref{Ypqmetric}), one can check that the corresponding four Killing 
vectors vanish at $\{y=1+\beta\}$, $\{\theta=\pi\}$, $\{y=1-\beta\}$, $\{\theta=0\}$, respectively. 
We denote the corresponding torus-invariant three-submanifolds of $\Ypq$ by $S_a$, 
$a=1,2,3,4$, respectively. As discussed at the end of section \ref{sec:obstruction},
the vectors in (\ref{Ypqv}) are normalized to be primitive, with signs fixed so that 
the minus complex structure pairs each vector with a radial vector that points inwards from the fixed point set.
This same toric data was derived via a  different route in appendix E of \cite{Couzens:2017nnr}. 
In this reference 
 the authors noted that if one writes $\v_a=(1,\w_a)$, with $\w_a\in\Z^2$, and plots the vectors $\w_a$ in the plane $\R^2\supset\Z^2$, the resulting 
so-called toric diagram is not convex. Correspondingly, for $q>p>0$ the vectors (\ref{Ypqv}) no longer define a convex polyhedral 
cone $\mathcal{C}$, via (\ref{Cv}). As discussed in section \ref{sec:toric}, this non-convexity is 
 a \emph{necessary} condition to admit a solution. Notice that when $p>q>0$ 
instead (\ref{Ypqv}) gives the convex toric data \cite{Martelli:2004wu} for the Calabi-Yau cones 
$C(Y^{p,q})$ over the Sasaki-Einstein five-manifolds $Y^{p,q}$ of \cite{Gauntlett:2004yd}, 
which of course do admit a compatible K\"ahler cone metric .

To understand the global structure more clearly, it is helpful to introduce a partial resolution $\hat{C}(\Ypq)$ of the cone 
$C(\Ypq)\cong \R_{>0}\times \Ypq$. Since $\Ypq$ may be viewed as a Lens space $S^3/\Z_p$ bundle over $S^2$, there is a 
natural choice: namely we take $\hat{C}(\Ypq)$ to be the total space of the associated $\C^2/\Z_p$ fibration over $S^2$, where $S^3/\Z_{\p}$ may be viewed as the boundary of $\C^2/\Z_{\p}$.  
When $p>q>0$ explicit Calabi-Yau metrics were constructed on these spaces in \cite{Martelli:2007pv}, and we may describe the 
 fibration structure following that reference. Denote standard complex coordinates on $\C^2$ by $(z_1,z_2)$, 
and let $U(1)_1$ rotate the $z_1$ coordinate with charge 1, and $U(1)_A$ be the anti-diagonal action in which $z_1$, $z_2$ have 
charges $-1$ and $+1$, respectively. Here the $\Z_{\p}$ action on $\C^2$ is via $\Z_{\p}\subset U(1)_A$. Then we twist the $\C^2/\Z_p$ fibre over $S^2$ using 
the canonical bundle $O(-2)$ of $S^2$ for the $U(1)_1$ action, and using $O(-\p-\q)$ for the action of $U(1)_A/\Z_\p$.  
The resulting space has a family of $\Z_{\p}$ orbifold singularities along the $S^2$ zero section. 
In particular notice that setting $z_1=0$ and $z_2=0$ both give $\C/\Z_p$ fibrations over $S^2$. The corresponding 
circle bundles $S^1\subset \C/\Z_p$ are easily computed using the above description: setting $z_2=0$ gives the circle bundle corresponding to 
$O(-p-q)$, while $z_1=0$ gives $O(-2p+(p+q))=O(-p+q)$. In fact this is precisely how the Chern numbers $p$ and $q$ 
are defined in the first place.

Since $\q>\p>0$ are arbitrary, it 
is convenient to temporarily set $\p=1$ in what follows, so that the above partial resolution is in fact a 
smooth manifold. The above description then identifies 
\bea
\hat{C}(\mathscr{Y}^{1,\q})  &=& \mbox{total space of\ } O(-1+\q)\oplus O(-1-q) \rightarrow S^2~.
\eea
Although this is a perfectly good smooth complex manifold, with an obvious 
holomorphic $U(1)^3$ action, it is \emph{not} a toric manifold in the usual sense. 
The problem can be seen by looking at the zero section of the second $O(-1-q)$ factor, which gives 
an embedded copy of 
\bea
M_{\q} &\equiv &  \mbox{total space of\ } O(-1+\q) \rightarrow S^2~.
\eea
For $\q>1$ this manifold is complex, but it has no regular non-constant holomorphic functions. 
Indeed, one can describe the holomorphic functions on $M_{\q}$ by Fourier decomposing 
along the fibre direction, or equivalently via charges under the $U(1)_1$ that acts on the fibre. 
A holomorphic function of charge $k\in\Z$ corresponds, in its dependence on the $S^2$ coordinates, to a section of $O(-k(-1+q))$. 
But for $k>0$ there are no such holomorphic sections, only meromorphic sections, and for $k<0$ the 
dependence of the function on the fibre coordinate $z_1$ is proportional to $z_1^k$, which has a pole singularity at the zero section $\{z_1=0\}\cong S^2$. 
When $k=0$ we have just the constant function. On the other hand, when $\q\leq 0$ instead $M_{\q}$ is a toric variety in the usual sense, admitting 
a compatible K\"ahler metric, and with a convex toric diagram.
 The spaces $\hat{C}(\mathscr{Y}^{1,\q})$, and more generally $\hat{C}(\mathscr{Y}^{\p,\q})$ for $\p\geq 1$, inherit this non-convexity 
and lack of holomorphic functions. In particular if one adds a point to the tip of the cones $C(\Ypq)=\R_{>0}\times \Ypq$, the resulting 
spaces are not affine varieties, since they are not generated by their ring of holomorphic functions. 
This is in contrast to the Calabi-Yau cones $C(Y^{p,q})=\R_{>0}\times Y^{p,q}$ for $\p>\q$.

The final geometric data that we need is the weights of the $U(1)^3$ action generated by (\ref{Ypqbasis}) 
on the fixed points of the partial resolution $\hat{C}(\mathscr{Y}^{\p,\q})$. The are two points which are 
fixed by the entire $U(1)^3$ action, namely the south and north poles of the $S^2$ zero section. 
Denoting the weights as  $\u_i^{(1)}$, $\u_i^{(2)}$, $i=1,2,3$, respectively, 
one finds 
\begin{align}
\label{Ypqu}
\p\, \u_1^{(1)} & =   (0,\p-\q,-\p+\q+1)~, \quad \p\, \u_2^{(1)} \ = \ (\p,\q,-1-\q)~, \quad  \p\, \u_3^{(1)} \ =\  (0,-\p,\p)~,\nn\\
\p\, \u_1^{(2)} & =   (\p,-\p,\p-1)~, \quad \ \, \p\, \u_2^{(2)} \ = \  (0,0,1)~, \quad  \p\, \u_3^{(2)} \ = \  (0,\p,-\p)~.
\end{align}
The normalizations here ensure that we have the correct corresponding weights for the torus action that enter the orbifold localization formula 
(\ref{localize}).  When $p>q>0$ the weights in (\ref{Ypqu}) are easily derived using toric geometry methods. In particular, 
$\u_1^{(1)}, \u_2^{(1)}, \u_1^{(2)}, \u_2^{(2)}$ are the outward-pointing generating edges of the polyhedral cone 
$\mathcal{C}$, and thus each has zero dot products with a pair of vectors $\v_a$ in (\ref{Ypqv}), and positive dot products 
with the remaining pair. On the other hand, $\u_3^{(1)}$, $\u_3^{(2)}$ are the weights on the tangent space 
of the blown up $S^2$, and these immediately follow from the basis (\ref{Ypqbasis}), given that $\partial_\phi$ rotates 
this $S^2$ with weights $\pm 1$ at the poles, while the other Killing vectors act trivially. This determines the weights 
at the fixed points in the basis of Killing vectors (\ref{Ypqbasis}) for $p>q$, but since the weights are linear 
in $p$ and $q$, in fact this then determines the weights for general $p$ and $q$.

Of course since the explicit supergravity solution is known in this case, one can impose the flux quantization conditions and compute the central 
charge in gravity directly, and this was done in \cite{Donos:2008ug}. 
However, let us see that we can instead recover this using \emph{only} the global complex geometry above, together 
with the general formulae in section \ref{sec:examplesT2}. Using (\ref{Ypqu}) and the localization formula (\ref{localize}) allows 
 us to compute
\begin{align}
& \int_{\Ypq} \eta \wedge \rho\wedge \rho \ = \ 4 (2\pi)^3\sum_{\{\mathrm{fixed\, points}\, p_A\, , A=1,2\}}\frac{1}{d_{p_A}}\prod_{i=1}^3 \frac{1}{(\vec{b},\u^{(A)}_i)}\label{Ypqconstraint}\\
& \qquad\quad= \frac{ 4 (2\pi)^3\, \p \left[b_3(\p-\q-2)\q + b_1 (\q -\p)\p + b_2(\q -\p)\q\right]}{b_3\left[(b_3(\p-1)+b_1 \p-b_2 \p\right]\left[b_3(\q+1)-b_1 \p -b_2 \q\right]\left[b_3(\p-\q-1)+b_2(\q-\p)\right]}~.\nn
\end{align}
Here $d_{p_A}=\p$ for both $A=1,2$, since these are the orders of the orbifold singularities at the two fixed points.  
We may then  impose the constraint (\ref{constraint5}), which sets (\ref{Ypqconstraint}) to zero. Solving this for $b_3$ gives
\bea\label{b3}
b_3 &=& \frac{(\q-\p)(b_1 \p + b_2\q)}{\q (\q-\p+2)}~,
\eea
while the holomorphic volume form $\Psi$ has charge $2$ under the R-symmetry vector when
\bea
b_1 &=& 2~.
\eea
This follows from (\ref{Ypqbasis}), together with the fact that $\Psi$ is only charged under $\partial_\psi$, with charge 1. 

To compute the central charge in (\ref{cSUGRAb31}) we may again use the localization formula (\ref{localize}). 
There are two natural choices of torus-invariant three-submanifolds, namely the copies of the Lens space fibres 
$S^3/\Z_\p$ over the south and north poles of the base $S^2$, where $\theta=\pi$ and $\theta=0$, respectively. Recall we denoted these 
by $S_2$, $S_4$, respectively.
 For $\p$ and $\q$ relatively prime, 
as in\cite{Gauntlett:2004yd} one can show $\Ypq\cong S^2\times S^3$, and 
$[S_2]=[S_4]=\p \in H_3(\Ypq,\Z)\cong \Z$. On the other hand, 
using (\ref{localize}) we compute
\bea
\int_{S_2} \eta\wedge\rho &=& 2 (2\pi)^2 \frac{1}{d_{p_1}}\prod_{i=1}^2 \frac{1}{(\vec{b},\u^{(1)}_i)} \ = \ \frac{2 (2\pi)^2\p}{b_3\left[b_3(\p-1)+b_1\p - b_2 \p\right]}~,\nn\\
\int_{S_4} \eta\wedge\rho &=& 2 (2\pi)^2 \frac{1}{d_{p_2}}\prod_{i=1}^2 \frac{1}{(\vec{b},\u^{(2)}_i)} \nn \\ & = & \frac{2 (2\pi)^2\p}{\left[b_3(\q+1)-b_1\p - b_2 \q\right]\left[b_3(\p-\q-1)+b_2(\q-\p)\right]}~.\label{fp3}
\eea
Notice that for general $\vec{b}=(b_1,b_2,b_3)$ these last two expressions are \emph{not} equal, as expected since although $S_2$, $S_4$ are in the same homology class, the 
forms that are being integrated are not closed. However, as shown in general in section \ref{sec:flux}, once we impose 
the constraint (\ref{constraint5}) these integrals become invariants of the homology class. We can see this very explicitly in this example: 
once we fix $b_3$ as in (\ref{b3}) so that the constraint (\ref{constraint5}) holds, we indeed find
\bea
\int_{S_2} \eta\wedge\rho \, = \, \int_{S_4} \eta\wedge\rho \  = \ \frac{2 (2\pi)^2\p \q^2(\q-\p+2)^2}{(\q-\p)(\p+\q)(b_1\p + b_2\q)\left[b_1(\q-\p+1)-b_2\q\right]}~.
\eea

We note that (\ref{Ypqconstraint}) may also be computed using the inward-pointing normals $\v_a$ in (\ref{Ypqv}) together with 
(\ref{toricconstraint}). For $p>q$, when we have a toric K\"ahler cone, 
these are guaranteed to be the same, as discussed in  section \ref{sec:toric}.  
For $q>p$ the fact these expressions agree supports our conjecture in section \ref{sec:toric}
that (\ref{toricconstraint}) also holds for non-convex toric 
geometries.\footnote{In later examples, discussed in sections
\ref{sec:Xpq}, \ref{sec:Zpq}, where we do not have explicit metrics or a simple explicit resolution of the 
relevant complex cones to use (\ref{Ypqconstraint}), we will utilize (\ref{toricconstraint}) to obtain some results which 
are then predicated on the validity of this conjecture.} Similarly, the expressions (\ref{fp3}) 
agree with (\ref{toric3}) using (\ref{Ypqv}) and setting the index $a$ in (\ref{toric3}) to $a=2$ and $a=4$, respectively. 

We now have everything that we need. The trial central charge (\ref{cSUGRAb31}) is
\bea
\cZ &=& \frac{12(2\pi)^2}{\int_{\sigma}\eta\wedge\rho}MN \ = \ \frac{12(2\pi)^2}{\frac{1}{\p}\int_{S_2}\eta\wedge\rho}MN~,
\eea
where $[\sigma]=1\in H_3(\Ypq,\Z)\cong \Z$ is the generating three-cycle. Setting $b_1=2$ and extremizing this function over $b_2$, we find the critical R-symmetry vector
\bea\label{criticalbYpq}
\vec{b} &=& (b_1,b_2,b_3) \ = \ \left(2, \frac{\p (\q-\p)}{\q},\frac{\p (\q-\p)}{\q}\right)~,
\eea
and central charge
\bea\label{cYpq}
 \cZ  |_\mathrm{on-shell} \ =  \ \csugra &=& \frac{6\p^2(\q-\p)(\p+\q)}{\q^2}MN~.
\eea
This agrees with the result for the explicit supergravity solution \cite{Donos:2008ug}. Moreover, using (\ref{criticalbYpq})
 and the definition of the basis (\ref{Ypqbasis}), one easily checks that
\bea
\sum_{i=1}^3 b_i \partial_{\varphi_i} \ = \ 2(\partial_\psi + \partial_z) \ = \ \xi~,
\eea
agreeing with the R-symmetry vector (\ref{Ypqxi}) of the original supergravity solution.

Notice that 
we have computed the central charge of  an AdS$_3$ solution using only the complex geometry of the cone as an input! Of course, this 
follows from the general prescription of section \ref{sec:general}. The supergravity solutions 
exist only for $\q>\p>0$ \cite{Donos:2008ug}. Note that  $N>0$ follows from 
the general formula (\ref{SsusyN}), since $A>0$ is the K\"ahler class of the $T^2$ and $S_{SUSY}>0$ from \eqref{Ssusy2}. 
Provided 
also $M>0$ then the central charge (\ref{cYpq}) is positive only if $q>p$. In supergravity recall that 
$\csugra$  is a positive multiple of the positive supersymmetric action 
$\Ssusy$ (\ref{cS}) and hence for consistency one must then take $M>0$. However, just looking at 
 (\ref{cYpq}) one could potentially have $q<p$ and $M<0$, and still have a positive central charge; 
the problem with this is that the $q<p$ complex cones are K\"ahler, and hence section
\ref{sec:obstruction} implies such a solution cannot exist.

In the full supergravity solution it
 is also possible to calculate the R-charges of baryonic operators in the SCFT that are associated with D3-branes wrapping supersymmetric three-cycles of $\Ypq\subset T^2\times \Ypq$.  
 As discussed at the end of section~\ref{sec:toric}, we may associate such an operator 
to each of the three-cycles $S_a\subset $, $a=1,2,3,4$, introduced above. The corresponding R-charges $R[S_a]$ in (\ref{rchgegenexpgeneral}) 
can be directly evaluated in this case \cite{Couzens:2017nnr} to give
\bea\label{YpqRgravity}
R[S_2]\ = \ R[S_4] \ =\ \frac{\ppp^2}{\qqq^2}N~,\qquad R[S_1]\ = \ R[S_3] \ = \ \frac{\qqq^2-\ppp^2}{\qqq^2}N~.
\eea
In particular, notice that (\ref{NR}) indeed holds, which supports the conjecture that (\ref{c1toric}) 
holds in general.
Since $N>0$, here one certainly needs $q>p$ in order that the R-charges of chiral primary operators are positive. 
It is interesting to note that in this case $R[S_2]+R[S_1]=N$.

\

\centerline{$*$}

\

We now switch gears and discuss the implications of the above results concerning the AdS$_3\times T^2\times \Ypq$ solutions 
in the context of the dual field theory proposed in \cite{Benini:2015bwz}. Specifically, 
\cite{Benini:2015bwz} proposed that these supergravity solutions are holographically dual to the 
four-dimensional $Y^{p,q}$ quiver gauge theories \cite{Benvenuti:2004dy}
compactified on $T^2$, with a baryonic twist. 
There are some reasons to hope for such an identification. For example, the fact that the $\Ypq$ is trivially fibred 
over the $T^2$ is consistent with the $Y^{p,q}$ quiver theory on $T^2$ with vanishing twist with respect to the flavour symmetry. Furthermore, the fact that $\Ypq$ has the same topology as $Y^{p,q}$, with a single three-cycle,
and that there is non-vanishing five-form flux on the product of this three-cycle with $T^2$ is consistent with the presence
of non-vanishing baryonic twist. Also, both geometries are specified by a pair of relatively prime integers $p,q$ and 
most strikingly, it was shown in  \cite{Benini:2015bwz} that the central charge as a function of $p,q$ obtained by 
$c$-extremization agrees with that for $\Ypq$.

However, as already noted in \cite{Couzens:2017nnr} there is an immediate problem with this identification: 
for the $Y^{p,q}$ field theories necessarily $p\geq q$ (for example, $p-q$ is the multiplicity of the $Z$ fields in the table below), 
while in the supergravity solutions based on $\Ypq$ instead $q>p$. Moreover, in this paper we have shown that 
if the complex cone $C(Y_5)$ is of Calabi-Yau type, then there is \emph{not} a corresponding 
AdS$_3\times T^2\times Y_5$ solution, with the same complex structure on $C(Y_5)$. 
Thus there cannot be a supergravity solution describing an RG flow across dimensions from AdS$_5\times Y^{p,q}$ in the UV to
AdS$_3\times T^2\times \Ypq$ in the IR, where the complex structure is preserved.\footnote{A flow from AdS$_5\times Y^{p,q}$ to AdS$_3\times T^2\times {\mathscr{Y}^{q,p}}$
is ruled out because the central charges would not agree.}

Nevertheless, it is interesting to re-examine the $c$-extremization calculation of  \cite{Benini:2015bwz}  in this case as it
exhibits a number of features that will recur in later examples. 
\begin{table}[h!]
\small{
\begin{center}
\begin{tabular}{|c|c|c|c|c|c|}
\hline
Field&Multiplicity&$R_0$-charge&$U(1)_{B}$& $U(1)_{F_{1}}$&$U(1)_{F_{2}}$\\
\hline\hline
$Y$&$(p+q) N^2$&$0$&$p-q$&$0$&$-1$\\
$Z$&$(p-q) N^2$&$0$&$p+q$&$0$&$1$\\
$U_1$&$p N^2$&$1$&$-p$&$1$&$0$\\
$U_2$&$p N^2$&$1$&$-p$&$-1$&$0$\\
$V_1$&$q N^2$&$1$&$q$&$1$&$1$\\
$V_2$&$q N^2$&$1$&$q$&$-1$&$1$\\
$\lambda$&$2p(N^2-1)$&1&0&0&0\\
\hline
\end{tabular}
\caption{The field content of the $Y^{p,q}$ quiver theories.}
\label{table1}\end{center}
}
\end{table}
The field content of the $Y^{p,q}$ quiver theories \cite{Benvenuti:2004dy} is summarized in the table \ref{table1}. The gauge group is $SU(N)^{2p}$, 
the $\lambda$ are the gauginos and the remaining fields are bifundamental matter fields. 
The $U(1)_{B}$ corresponds to the baryonic symmetry associated to the single three-cycle
$\sigma=1\in H_3(Y^{p,q},\Z)\cong \Z$, while $U(1)_{F_i}$, $i=1,2$, are flavour symmetries corresponding to $U(1)$ isometries 
under which the holomorphic volume form is uncharged. In particular $U(1)_1\subset SU(2)$ is the Cartan of the 
$SU(2)$ isometry that acts on the round $S^2$ in the metric. Note that $R_0$ is {\it not} the R-charge of the dual 
SCFT in $d=4$ (which can be found in \cite{Benvenuti:2004dy}). Instead $R_0$ is a simple assignment of a fiducial R-charge 
that can be used in the $c$-extremization procedure for the putative $d=2$ SCFT; we can use any assignment
for $R_0$ compatible with the usual requirements of an R-symmetry (every term in the superpotential has $R_0$-charge 2 and the gauginos, $\lambda$, have $R_0$-charge 1).

We consider these $d=4$ SCFTs theories wrapped on $T^2$, introducing only baryonic flux in the topological twist:
\bea\label{Ypqtop}
T_{\mathrm{top}} & =&  \betabeta T_{B}~.
\eea
Geometrically, if these were to be dual to solutions of the form AdS$_3\times Y_7$, then this twisting would
be associated with a product $Y_7=T^2\times Y_5$, as noted above. The trial R-charge is a linear combination
\bea\label{YpqtrialR}
T_{\text{trial}} &=&  T_{R_0}+\zeta T_{B}+ \epsilon_{1} T_{F_{1}}+\epsilon_{2} T_{F_{2}}~,
\eea
where $\zeta$, $\epsilon_i$ are parameters. We now substitute this data into the trial central charge 
given in \cite{Benini:2012cz} and extremize over the parameters $\eta$, $\epsilon_i$. The extremal point
has
\begin{align}\label{exzetepypq}
\zeta \ = \ \frac{q^2-p^2}{pq^2}~,\qquad \epsilon_1 \ = \ 0~,\qquad \epsilon_2\ =\ \frac{p^2-q^2}{pq}~.
\end{align}
and one finds \cite{Benini:2015bwz}
\bea\label{cextYpq}
c_{\text{c-ext}} & =& 6\frac{\p^2(\q-\p)(\p+\q)}{\q^2}\betabeta N^2~.
\eea
This formally agrees with the geometric result (\ref{cYpq})
on making the identification
\bea\label{Mbeta}
M &=&  N \betabeta~,
\eea
of geometric and field theory baryonic flux parameters $M$, $\betabeta$, respectively. 
Physically, (\ref{Mbeta}) is consistent with the fact that $M$ is the number of units of $F_5$ flux 
through $T^2\times \sigma$, where $\sigma$ is the three-cycle in $Y_5=Y^{p,q}\cong S^2\times S^3$ 
that generates the $U(1)_B$ symmetry. 

Of course, the immediate problem with identifying the field theory result 
(\ref{cextYpq}) with the gravity result (\ref{cYpq}) is that the ranges of $p$ and $q$ 
are complimentary. However, as discussed above, the central charge 
(\ref{cextYpq}) can be positive if $p>q$ and also $\betabeta<0$. However, this may be ruled out 
in \emph{field theory} by examining the R-charges of fields. These are easily computed using \eqref{exzetepypq} and we get
\bea\label{YpqR}
R_{\text{c-ext}}[U_1] \ = \ R_{\text{c-ext}}[U_2] \ = \ \frac{p^2}{q^2}N~, \qquad R_{\text{c-ext}}[Z] \ = \ R_{\text{c-ext}}[Y] \ = \ \frac{q^2-p^2}{q^2}N~.
\eea
The gauge-invariant baryonic operators constructed from the fields $Z, U_1, Y, U_2$ precisely 
correspond to D3-branes wrapped on the three-cycles $S_1,S_2,S_3,S_4$, respectively \cite{Benvenuti:2004dy}. 
Thus (\ref{YpqR}) also match the supergravity results (\ref{YpqRgravity}), except 
again the ranges of $p$ and $q$ are complimentary. Moreover, from 
the field theory results (\ref{YpqR}) where $p>q$ one sees that 
$R[Z]$ and $R[Y]$ are \emph{negative}, which is a contradiction  for a chiral
operator. The conclusion must be that such a superconformal fixed point, with such R-charges, does not exist.
We discuss possible refinements of the $c$-extremization procedure in section
\ref{sec:discuss}.

We shall give some more elaborate examples of a similar ``matching'' in the remainder of this section. 
From these examples it seems clear that \emph{mathematically} the central charges and R-charges being computed on both 
sides will \emph{always agree}. Notice that the quiver theory has encoded in it the complex cone geometry, which 
arises as the mesonic moduli space, and thus this same complex data is present in both descriptions. 
It must then be the case that the supergravity formulae of section \ref{sec:general} are computing the 
\emph{same} geometric objects as $c$-extremization in the field theory. This is currently far from obvious, but 
we state it here as a conjecture, based on examples.
However, one must be careful when interpreting the results physically, 
as we have seen. There can be obstructions to the existence of supergravity solutions/superconformal fixed points 
with the assumed properties. The current status is that the field theory dual to the 
AdS$_3\times T^2\times \Ypq$ solutions are not known! On the other hand, the fate of the $Y^{p,q}$ quiver 
gauge theories on $T^2$, with only baryonic twist, is also unknown!

To conclude this subsection we briefly consider the analogous results for the case of $T^{1,1}$.
From the field theory perspective we are interested in the reduction of the Klebanov-Witten theory \cite{Klebanov:1998hh}
on $T^2$ with baryon flux only. The four-dimensional theory has four bifundamental fields transforming under an $SU(N)^2$ gauge group with two abelian non-R flavour symmetries and one baryonic symmetry. The charge assignments of the fields may be obtained by setting $p=1$ and $q=0$ in table \ref{table1}. After carrying out the c-extremization, it is easy to see that the trial central charge does not admit a critical point. The situation is therefore, in some sense, worse than that of $Y^{p,q}$ field theories as the putative theory does not admit a critical point regardless of the positivity of the R-charges. On the gravity side, we have already shown that there are no AdS$_3\times T^2\times T^{1,1}$ type solutions with the same complex structure on the conifold. Again, the situation is worse than the case of $Y^{p,q}$ since we can't even satisfy the constraint equation
\eqref{constraint5}. Indeed, after inputting the $d=4$ toric vectors for the conifold into the constraint equation \eqref{toricconstraint}, which gives the same result as \eqref{Ypqconstraint} after setting $p=1,q=0$, we find that the only solution is $b_1=0$, contradicting $b_1=2$. 

\subsection{$Y_5=\Labc$}\label{sec:Labc}
In this section we analyse a class of supersymmetric AdS$_3$ solutions, labelled $Y_5=\Labc$, that are
specified by three positive integers, $\mathtt{a},\mathtt{b},\mathtt{c}$, satisfying some additional conditions given below. Just as the $\Ypq$ solutions are analogues
of the $Y^{p,q}$ Sasaki-Einstein metrics \cite{Gauntlett:2004yd}, the $\Labc$ solutions are analogues\footnote{The reason for using different fonts for the integers, {\it e.g.} $\mathtt{a}$ versus $a$, will become clear later.}
 of the $L^{a,b,c}$ Sasaki-Einstein
metrics constructed in \cite{Cvetic:2005ft}. Moreover, the family of $\Labc$ solutions include the
$~\Ypq$ solutions discussed in the last subsection as a special case.

The local metrics for $Y_5=\Labc$ were first constructed in Appendix C of \cite{Donos:2008hd}. 
Here we present sufficient conditions for these to describe regular geometries with suitably quantized flux.
Following this analysis we are able to calculate the central charge and R-charges for certain baryonic operators directly.
Using coordinates similar to \cite{Martelli:2005wy},
the six-dimensional local K\"ahler metric is given 
by\footnote{Start from (C.4), (C.5)  and (C.27) of \cite{Donos:2008hd}, with $Q=0$ ({\it i.e.} vanishing three-form flux) and
then write $\mu_1=\cos\frac{\theta}{2}$, $\mu_2=\sin\frac{\theta}{2}$. After identifying
$w=-4 \xi$, $
\cos \theta= \frac{1}{q_{2}-q_{1}}(8 \eta -(q_{1}+q_{2}))$, $\phi_{1}=\frac{1}{8}(4 \phi+ q_{2} \psi)$, $\phi_{2}=\frac{1}{8}(4 \phi + q_{1} \psi)$, $q_{1}= 4$, $q_{2}=4 \beta$, and $\lambda=-(1+\alpha)$ one obtains precisely \eqref{loclalabcsix} and \eqref{LabcFG}. \label{footnote:coordchange}}
\bea\label{loclalabcsix}
\dd s^{2}_6 & = 
 & \frac{\eta-\xi}{\mathcal{F}(\xi)}\dd \xi^{2}+\frac{\mathcal{F}(\xi)}{\eta-\xi}(\dd \phi +\eta \dd \psi)^{2}+\frac{\eta-\xi}{\mathcal{G}(\eta)}\dd \eta^{2}+\frac{\mathcal{G}(\eta)}{\eta-\xi}(\dd \phi +\xi \dd \psi)^{2}\nn\\
&& + \, \dd s^2(T^2)~.
\eea
This is fixed by two functions, which are explicitly given by
\begin{align}\label{LabcFG}
\mathcal{F}(\xi) \ = \  -\mathcal{G}(\xi)-(1+\alpha) \xi^{2}~,\qquad
\mathcal{G}(\eta) \ =\  - (\eta-1)(\eta-\beta)~,
\end{align}
where $\alpha,\beta$ are constants.
The type IIB AdS$_3$ solution with $Y_7=T^2\times Y_5$ can be obtained from \eqref{ansatz}, \eqref{metric} using the data
\begin{align}
\ex^B& \ = \ \frac{1+\alpha}{4 (\eta-\xi)}~,\qquad
P \ = \ \frac{(1+\alpha)\xi}{\eta-\xi}(\dd \phi+\eta \dd \psi) {+ \dd \phi + \frac{\beta+1}{2} \dd \psi}~,\nn\\
F &{  \, \ = \  -\frac{2}{(1+\alpha)} \left( \dd z +\dd \phi +\frac{\beta+1}{2} \dd \psi \right) \wedge(\dd \eta-\dd \xi)+ {2} \dd \phi \wedge \dd \eta
-  2\vol(T^{2})}~.
\end{align}
In order to get a positive definite metric we take $\alpha>0$ and $\beta>1$ and the ranges of the coordinates $\eta$ and $\xi$ are
taken to be $1\le\eta\le\beta$ and $\xi_-\le\xi\le \xi_+$, where $\xi_\pm$ are the roots of the quadratic $\mathcal{F}$.
Note that we have $\xi_-<0<\xi_+<1$  
and therefore $\eta-\xi>0$ everywhere. 

We next analyse the additional conditions that are required in order to have a well behaved metric on a
regular manifold $Y_7=T^2\times Y_5$. The $T^2$ factor is not relevant and so we can concentrate on $Y_5$, closely following
the analysis for the Sasaki-Einstein metrics $L^{a,b,c}$ given in \cite{Cvetic:2005ft}. 
The metric on $Y_5$ has three commuting Killing vectors given by $\partial_{\psi},~\partial_{\phi}$ and the R-symmetry Killing vector $\partial_{z}$. The complete metrics will be cohomogeneity two, with $U(1)^3$ principal  orbits. These orbits 
degenerate at the roots of the two quadratic functions $\mathcal{F}(\xi)$ and $\mathcal{G}(\eta)$, and there are four such degeneration surfaces. Specifically, we find that the degenerating Killing vectors at the surfaces
$\eta=1,\beta$ and $\xi=\xi_{\pm}$ are given by 
\begin{align}\label{Killing1new}
k_{1}& \ = \ -\frac{2}{\beta-1}\left( \frac{\partial}{\partial \psi}-\frac{\partial}{\partial \phi}- 
\frac{\beta-1}{2} \frac{\partial}{\partial z}\right)~,\nn\\
k_{\beta}& \ =\ \frac{2}{\beta-1}\left( \frac{\partial}{\partial \psi}-\beta\frac{\partial}{\partial \phi}+ 
\frac{\beta-1}{2} \frac{\partial}{\partial z}\right)~,\nn\\
l_{\pm}& \ = \ \mp\frac{2}{\alpha(\xi_{+}-\xi_{-})}\left( \frac{\partial}{\partial \psi}-\xi_{\pm}\frac{\partial}{\partial \phi}-\left(\alpha \xi_{\pm} +\frac{\beta+1}{2}\right) \frac{\partial}{\partial z}\right)~,
\end{align}
respectively. These vectors have been normalized to have unit surface gravity and the overall signs have been chosen for convenience.

To ensure that the collapsing orbits extend smoothly onto the degeneration surfaces without generating conical singularities, we need to impose suitable conditions on the parameters $\alpha,\beta$. After some further analysis
we find that the regular solutions, labelled $Y_5=\Labc$, are fixed by 
four positive integers $\aaaa$, $\bbbb$, $\cccc$ and $\dddd$, satisfying the following conditions 
\begin{align}\label{frakintrel1}
-\aaaa+\bbbb& \ = \ \cccc+\dddd~,\nn\\
\bbbb > \cccc\ge  \dddd>0~,&~\qquad  \bbbb>\aaaa>0~.
\end{align}
In particular, only three of these integers are independent and the solutions can be labelled by $\aaaa$, $\bbbb$, $\cccc$.
We further demand that $\mathrm{hcf}(\aaaa,\bbbb,\cccc,\dddd)=1$, which implies that any three integers are coprime. In addition we demand that $\aaaa,\bbbb$ are each coprime to each of 
$\cccc,\dddd$. In terms of these integers, the parameters $\alpha$, $\beta$ are given by\footnote{\label{bip}Note that when 
$\cccc=\dddd$, the two roots of $\eta$ coincide. This case, which leads back to the $\Ypq$ examples, needs to be treated with different coordinates; for example the change of coordinates described in footnote \ref{footnote:coordchange} is no longer valid.}
\begin{align}
\alpha \ = \ \frac{\aaaa \bbbb}{\cccc \dddd}~,\qquad
\beta \ = \ \frac{\cccc}{\dddd}~.
\end{align}
The roots of $\mathcal{F}(\xi)$ are then explicitly given by
\begin{align}
\xi_{+} \ =  \ \frac{\cccc}{\bbbb}~,\qquad
\xi_{-}\ =\ - \frac{\cccc}{\aaaa}~.
\end{align}
Furthermore, in terms of $\aaaa$, $\bbbb$, $\cccc$, $\dddd$ the linear relation between the four degenerating Killing vectors
\eqref{Killing1new} is given by
 \begin{align}
\cccc k_1 +\dddd k_\beta +\aaaa l_{-}  - \bbbb l_{+} \ = \ 0~.
 \end{align}
 
Now, since $\mathrm{hcf}(\bbbb,\cccc)=1$, B\'ezout's identity implies there exist non-unique $\mathtt{k}, \mathtt{l} \in \Z$,
satisfying 
\be
\bbbb \mathtt{l }+\cccc \mathtt{k} \ = \ 1~.
\ee
Let us fix such a pair $\mathtt{k}, \mathtt{l}$. Using this we can replace the coordinates $(\phi,\psi,z)$
parametrizing the three-torus 
with a new set of coordinates, $(\etpsi_1,\etpsi_2,\etpsi_3)$, via
\begin{align}\label{epsiecds}
\phi& \ = \ -\frac{2 \cccc}{\cccc-\dddd} \etpsi_{1}
+\frac{ 2 \cccc [ \aaaa(\aaaa+\bbbb)+\dddd(\cccc-\dddd)]}{\aaaa (\aaaa+\bbbb)(\cccc-\dddd)}\etpsi_{2}\nn\\
&\qquad +\frac{2[ \cccc \dddd (\cccc-\dddd)\mathtt{k} +(\aaaa+\bbbb)\bbbb (\dddd \mathtt{l} + \cccc (\mathtt{k}+\mathtt{l}))]}{\bbbb (\aaaa+\bbbb)(\cccc-\dddd)}\etpsi_{3}~,\nn\\
\psi& \ = \  \frac{2 \dddd}{\cccc-\dddd} \etpsi_{1}- \frac{4 \dddd (\aaaa+\dddd)}{(\aaaa+\bbbb)(\cccc-\dddd)}\etpsi_{2}-\frac{4 \dddd[ \aaaa+\bbbb + \aaaa \mathtt{k} (\bbbb-\cccc)]}{\bbbb (\aaaa+\bbbb)(\cccc-\dddd)}\etpsi_{3}~,\nn\\
z&\  = \ \etpsi_{1}~.
\end{align}
The Killing vectors $\partial_{\etpsi_i}$ generate an effective torus action, 
and moreover the range of these coordinates is $0\le\etpsi_i\le 2\pi$. Specifically, we
have
\begin{align}\label{vecetorus}
\begin{pmatrix}
k_{\beta}\\
k_{1}\\
l_{+}\\
l_{-}
\end{pmatrix} \ = \  
\begin{pmatrix}
1&0&0\\
1&-\aaaa \mathtt{k} & \bbbb\\
1&\aaaa \mathtt{l} &\cccc\\
1& 1&0
\end{pmatrix}
\begin{pmatrix}
\partial_{\etpsi_1}\\
\partial_{\etpsi_2}\\
\partial_{\etpsi_3}
\end{pmatrix}~.
\end{align}
As in the discussion around \eqref{Ypqv} we can extract the toric data 
on the cone over $\Labc$. Specifically, 
the four vectors that define vanishing $U(1)$s along complex codimension one submanifolds are given by 
\begin{align}\label{Labcv}
\v_1  \ = \ (1,1,0)~, \quad \v_2 \ = \ (1,-\aaaa \mathtt{k},\bbbb)~, \quad \v_3 \ = \ (1,\aaaa \mathtt{l},\cccc)~, \quad \v_4  \ = \  (1,0,0)~, 
\end{align}
with the corresponding submanifolds being defined by $\xi=\xi_{-},~\eta=1, ~\xi=\xi_{+},~\eta=\beta$, respectively.
We denote the corresponding torus-invariant three-submanifolds of $\Labc$ by $S_a$, 
$a=1,2,3,4$, respectively. 
One can compare this with the toric data for the $L^{a,b,c}$ Sasaki-Einstein metrics given in eq. (3.2) of 
\cite{Franco:2005sm}, and we observe that the sign of the second entry in $\v_2$ and $\v_3$ differs. This implies that the 
putative toric diagram that we can construct for the complex geometry on the cone associated with the $\Labc$ solutions is not convex (in contrast  to that for $L^{a,b,c}$).

We can recover the $\Ypq$ solutions as a special case of the $\Labc$ solutions. 
Specifically,  we should set
\be\label{recoverypq}
\bbbb=\ppp+\qqq~,~~\aaaa=\qqq-\ppp~,~~\cccc=\dddd=\ppp~,
\ee
which solves $-\aaaa+\bbbb=\cccc+\dddd$ (and we recall the comment in footnote \ref{bip}). 
In particular, if one substitutes this into \eqref{Labcv}, and carries out a suitable 
$SL(3,\mathbb{Z})$ transformation, then we recover \eqref{Ypqv}.

We now discuss the quantization of the five-form flux for the $\Labc$ solutions. The relevant part of the five-form in \eqref{ansatz}, tangent to $Y_7=T^2\times \Labc$,
is given by
\begin{align}
L^{-4}F_5|_{Y_7}&\ = \  -\left. \frac{1+\alpha}{4(\eta-\xi)^{2}}\right(  \vol(T^{2})\wedge (\mathcal{F}(\xi)\dd \eta +\mathcal{G}(\eta)\dd \xi)\wedge \dd \phi \wedge \dd \psi \nonumber\\
& \qquad + \vol(T^{2})\wedge(\dd z+P)\wedge [\xi \dd \eta \wedge (\dd \phi+\xi \dd \psi)-\eta \dd \xi \wedge (\dd \phi+\eta \dd \psi)]\nn\\
&\qquad + (\eta-\xi)^{2} (\dd z+P)\wedge \dd \eta \wedge \dd \xi \wedge \dd \phi \wedge \dd \psi  \left. \frac{}{}\right)\,.
\end{align}
The topology of $\Labc$ is $S^2\times S^3$ and can be established as in \cite{Cvetic:2005ft,Franco:2005sm}. 
Thus, there is a single generator, $\sigma$, of $H_3(\Labc,\mathbb{Z})\cong\Z$. With respect to the torus invariant
three manifolds $S_a$, defined above, we have the homology relations in $H_3(\Labc,\Z)$:
\begin{align}
[S_{1}]\ = \ \aaaa\sigma~,\qquad
{[S_{2}]}\ = \ \cccc\sigma~,\qquad 
[S_{3}]\ = \ \bbbb\sigma~,\qquad 
[S_{4}] \ = \ \dddd \sigma~.
\end{align}
We find that it is sufficient if we impose the following quantization conditions 
\begin{align}
\frac{L^4}{\pi l_s ^4 g_s}\ =\  \frac{  \aaaa^2 \bbbb^2 \dddd}{ (\aaaa\bbbb+\cccc \dddd)^2}N~,\qquad\qquad
\frac{1}{4\pi}\vol(T^{2}) \ = \ \frac{(\aaaa\bbbb +\cccc \dddd) \cccc}{\aaaa \bbbb }\frac{M}{N}~,\label{Baryoniccharge}
\end{align}
where $N,M$ are positive integers. Indeed we then find
\begin{align}\label{quantfqlabc}
\frac{1}{(2 \pi l_s)^{4}g_s}\int_{Y_{5}} F_5 \ = \ N~, \qquad\quad
\frac{1}{(2 \pi l_s)^{4}g_s}\int_{T^2\times \sigma} F_5& \ = \ M~.
\end{align}

It is now straightforward to calculate the central charge for the dual two-dimensional CFT using \eqref{cSUGRAgen}, and
we obtain 
\be\label{faswqw}
\csugra \ = \ 6  \frac{\aaaa \bbbb \cccc \dddd }{\aaaa\bbbb +\cccc \dddd}N M~.
\ee
It is also possible to calculate the R-charges of baryonic operators in the CFT that are associated with D3-branes wrapping supersymmetric three-cycles of $T^2\times \Labc$. As discussed in \cite{Couzens:2017nnr}, these are precisely the 
three-cycles $S_a$ discussed above. 
Using the general expression \eqref{rchgegenexpgeneral} and the
first expression in \eqref{quantfqlabc} we find\footnote{In carrying out the explicit integrals,
one needs to use a good set of coordinates on the three-cycle. The coordinates introduced in \eqref{epsiecds} have the feature that
on the surface $\eta=\beta$, it is the Killing vector $\partial_{\etpsi_{1}}$ that is degenerating. Thus, we can use the coordinates 
$\xi,\etpsi_2,\etpsi_3$ to parametrize $S_\beta$ and the integral is straightforward to carry out. For each of the
other three degenerating surfaces we should use another set of coordinates, obtained by using a suitable $SL(3,\mathbb{Z})$ transformation on the $\partial_{\etpsi_i}$ in order to have a similar feature.}
\begin{align}\label{areare}
R[S_{2}]& \ = \ R[S_{4}]\ =\  \frac{\cccc \dddd }{\aaaa\bbbb+\cccc\dddd}N~,
\nn\\
R[S_{1}]& \ = \ R[S_{3}] \ = \ \frac{ \aaaa \bbbb}{\aaaa\bbbb +\cccc \dddd }N ~.
\end{align}
We observe that once again (\ref{NR}) indeed holds  which provides further support for the conjecture that (\ref{c1toric}) 
holds in general. It is interesting to note that in this case we also have $R[S_2]+R[S_1]=N$.
We also
notice that upon substituting the $\Ypq$ values for $\aaaa,\bbbb,\cccc,\dddd$, 
given in \eqref{recoverypq}, into \eqref{faswqw} and \eqref{areare} we recover \eqref{cYpq} and \eqref{YpqRgravity}, respectively.

In principle, we can also obtain the central charge using the results of section 3. To use the formula
\eqref{localize} we would need to find a suitable resolution of the cone over $\Labc$. We will not pursue that here, but instead we 
point out that we obtain the correct central charge, $\cZ  |_\mathrm{on-shell} = \csugra$, if we again use  \eqref{toric3} and \eqref{toricconstraint}, each of which 
may be checked explicitly in this case. 
In other words, this provides further evidence for the conjecture we made in section \ref{sec:toric}.

\

\centerline{$*$}

\

We now consider the $d=4$ quiver gauge theories dual to the AdS$_5\times L^{a,b,c}$ type IIB solutions, and reduce
them on a $T^2$ with non-vanishing baryonic flux. Assuming that these field theories flow in the far IR to a $d=2$ $(0,2)$ SCFT
we can attempt to calculate the putative central charge, $c_{\text{c-ext}}$, as well as the R-charges of
certain operators, using $c$-extremization \cite{Benini:2012cz,Benini:2013cda,Benini:2015bwz}. 

The field content of the $L^{a,b,c}$ quiver theories, which have gauge group $SU(N)^{a+b}$ \cite{Benvenuti:2005ja,Butti:2005sw,Franco:2005sm} is summarized in the table \ref{table2}. Once again  the $\lambda$ are the gauginos and the remaining fields are bifundamental or adjoint matter fields.
\begin{table}[h!]
\small{
\begin{center}
\begin{tabular}{|c|c|c|c|c|c|}
\hline
Field&Multiplicity&$R_0$-charge&$U(1)_{B}$&$U(1)_{F_{1}}$&$U(1)_{F_{2}}$\\
\hline\hline
$Y$&$bN^2$&$0$&$a$&$1$&$0$\\
$Z$&$aN^2$&$0$&$b$&$0$&$k$\\
$U_{1}$&$dN^2$&$1$&$-c$&$0$&$l$\\
$U_{2}$&$cN^2$&$1$&$-d$&$-1$&$-l-k$\\
$V_{1}$&$(b-c)N^2$&$1$&$b-d$&$-1$&$-l$\\
$V_{2}$&$(b-d)N^2$&$1$&$b-c$&$0$&$l+k$\\
$\lambda$&$(a+b)(N^2-1)$&1&0&0&0\\
\hline
\end{tabular}
\caption{The field content of the $L^{a,b,c}$ quiver theories.}
\label{table2}
\end{center}
}
\end{table}

Here, $a,b,c,d$ satisfy
\begin{align}\label{frakintrel}
a+b \ = \ c+d~,\nn\\
b\ge a, c, d>0~,
\end{align}
and $k,l\in\mathbb{Z}$ are chosen to satisfy 
\be\label{blckcond}
b l+ c k  \ = \ 1~.
\ee
Once again $R_0$ is a fiducial choice of R-charge to use in $c$-extremization. 
Note that the integers $k,l$ are not unique: for a given $(k,l)$ satisfying \eqref{blckcond} we also have that $(k+nb,l-nc)$ 
satisfies \eqref{blckcond} for arbitrary $n\in\mathbb{Z}$. 
In the quiver gauge theory, this ambiguity is associated with the following redefinition of
the flavour symmetry: $U(1)_{F_2}\to U(1)_{F_2}+n\, U(1)_{B}-na\, U(1)_{F_1}$.

We now wrap these quiver theories on a $T^2$ and topologically twist with baryonic flux only, as in \eqref{Ypqtop}. The
trial R-symmetry is constructed as in \eqref{YpqtrialR}. 
After substituting this into the trial central charge given in \cite{Benini:2012cz} and extremizing over the parameters $\eta$, $\epsilon_i$ we find the central charge 
\be
c_{\text{c-ext}}(L^{a,b,c}) \ = \ 6\frac{ a b c d}{(b-c)(b-d)}\betabeta N^2 \ = \  6\frac{a b c d}{a b -cd}\betabeta  N^2~.
\ee
The middle expression is manifestly positive for $\betabeta>0$ by \eqref{frakintrel}. Furthermore\footnote{Observe that setting $c=d=p$, $a=p-q$, $b=p+q$, which takes $L^{a,b,c}$ to $Y^{p,q}$, we precisely recover
the results in \eqref{cextYpq} and \eqref{YpqR}.}, the R-charges are 
\begin{align}
R_{\text{c-ext}}[Y]& \ = \ R_{\text{c-ext}}[Z] \ = \  -\frac{ab}{(b-d)(b-c)}N \ = \  \frac{ab}{ab-cd}N~,\nn\\
~R_{\text{c-ext}}[U_1]& \ = \ R_{\text{c-ext}}[U_2] \ =\ \frac{ c d}{(b-d)(c-b)}N~ \ = \ -\frac{cd}{ab-cd} N~.
\end{align}
From the R-charges it is clear that it is not possible to suitably tune the integers to simultaneously make all R-charges positive as required for chiral operators. The conclusion is the same as that of the previous section: there is no 
such superconformal fixed point. We make some additional comments on this point in section
\ref{sec:discuss}.

Nevertheless, it is interesting to compare the result of $c_{\text{c-ext}}(L^{a,b,c})$ and $R_{\text{c-ext}}$ 
with the results obtained for $\Labc$ in \eqref{faswqw}. A na\"ive matching as in the $Y^{p,q}$ versus $\Ypq$ scenario does not
work. However, if we make the formal identification
\be
a\ = \ -\aaaa~.
\ee 
then they do precisely agree, as do the defining relations for the integers. 
Of course this matching is only formal, since both $a$ and $\aaaa$ must be positive.

It remains an open question to identify the CFTs dual to the explicit AdS$_3\times T^2\times \Labc$ solutions.
It also remains an open question to determine the IR behaviour of the $L^{a,b,c}$ quiver gauge theories
reduced on $T^2$, twisted with baryonic flux.

\subsection{$Y_5=X^{p,q}$}\label{sec:Xpq}

The examples in this subsection and the next are somewhat different. We have already argued in section \ref{sec:obstruction} that if the complex manifold $C(Y_{5})=\R_{>0} \times Y_5$ 
is of Calabi-Yau type, then there is no corresponding AdS$_3\times T^2\times Y_5$ solution. On the other hand, we have also seen in section \ref{sec:Ypq} that if one formally applies  the geometric formulae (\ref{toric3}), (\ref{toricconstraint}) to the Calabi-Yau cones $C(Y^{p,q})$ with $p>q$,  the resulting central charge agrees with 
that computed using $c$-extremization for the four-dimensional $Y^{p,q}$ quiver gauge theories \cite{Benvenuti:2004dy} compactified on $T^2$. Similar remarks apply 
to the  $L^{a,b,c}$ quiver gauge theories \cite{Benvenuti:2005ja,Butti:2005sw,Franco:2005sm} and section \ref{sec:Labc}.
In the next two subsections we show that this formal matching continues to hold 
for other complex cones $C(Y_{5})$ of Calabi-Yau type. In particular for the examples in this subsection $b_3(Y_5)=2$, for which we must also impose (\ref{ratios}). 
In this case there is no extremization to perform in the geometric computation, since flux quantization determines uniquely the R-symmetry vector. 
The result perfectly matches the $c$-extremization result for the $X^{p,q}$ quiver gauge theories \cite{Hanany:2005hq} compactified on $T^2$.
In this subsection we will first discuss these points before concluding with a conjecture concerning
the existence of a new family of AdS$_3\times T^2\times \mathscr{X}^{p,q}$ solutions, with $q>p>0$.

The $X^{p,q}$ Calabi-Yau cones are toric, with inward-pointing normal vectors 
\bea\label{Xpqv}
\v_1 &=& (1,1,0)~, \quad \v_2 \ = \ (1,2,0)~, \quad \v_3 \ = \ (1,1,p)~, \nn\\ 
\v_4 & = & (1,0,p-q+1)~, \quad \v_5 \ = \ (1,0,p-q)~.
\eea
Here $p> q>0$ are integers. Being Calabi-Yau cones, these give rise to supersymmetric AdS$_5\times X^{p,q}$ solutions 
of type IIB which are holographically dual to 
 corresponding four-dimensional $\mathcal{N}=1$ quiver gauge theories. These gauge theories were presented
in \cite{Hanany:2005hq}, and by construction may be 
Higgsed to the $Y^{p,q}$ theories. In particular for $p=2$, $q=1$ the complex cone is the canonical complex cone over the second del Pezzo surface, dP$_2$. 
The five-manifolds $Y_5=X^{p,q}$ have $b_3(X^{p,q})=2$. 

Using this data 
we may apply our general formulae from section \ref{sec:examplesT2}. 
Inserting (\ref{Xpqv}) into the constraint equation (\ref{constraint5}) using (\ref{toricconstraint}) immediately gives
\bea
0 =   \frac{b_{1}[b_{3}^2 -2 b_{3} p (b_{1}+b_{2}(q-1)) -p(p-q) b_{2}(b_{1}(1+p-q)+b_{2}(q-1))]}{b_{2}b_{3}[b_{3}+(-2 b_{1}+b_{2})p][b_{3}-(b_{1}-b_{2})(p-q)][b_{1}(1+p-q)+b_{2}(q-1)-b_{3}]}
\eea
This may be solved for $b_3$, giving
\bea
b_{3} & =& p(b_{1} -b_{2})+b_{2} p q 
\nn\\
&& 
+\sqrt{p[b_{1}^2 p+b_{2}^2(p-1)(q-1)q+b_{1}b_{2}(p(p-1)+q(q-1))]}~.
\eea
There are five toric divisors, which map to the facets $\{(\vec{y},\v_a)=0\}$ in the moment map cone $\mathcal{C}$, $a=1,\ldots, 5$. 
These toric divisors are in turn cones over  torus-invariant  three-dimensional submanifolds $S_a$. 
On the other hand, since $b_2(X^{p,q})=2$ there are two generating three-cycles 
$\sigma_1$, $\sigma_2$. The homology relations in $H_3(X^{p,q},\Z)$ are correspondingly
\bea\label{Xpqhom}
[S_1] & = & \sigma_1~, \quad [S_2] \ = \ \sigma_2~, \quad [S_3] \ = \  -\sigma_{1}-2 \sigma_{2}~, \quad [S_4] \ = \  p \sigma_{1}+(p+q) \sigma_{2}~,\nonumber  \\
\left[ S_5 \right] & = & -p \sigma_{1}+(1-p-q) \sigma_{2}~.
\eea
Flux quantization (\ref{aM}) imposes
\bea
\int_{\sigma_{1}} \eta \wedge \rho & = & \frac{2(2 \pi \ell_s )^4 g_s}{A L^4}M_{1}~,\qquad \int_{\sigma_{2}} \eta \wedge \rho \ = \  \frac{2(2 \pi \ell_s )^4 g_s}{A L^4}M_{2}~, 
\eea
which implies
\bea
\frac{\int_{\sigma_{1}} \eta \wedge \rho }{\int_{\sigma_{2}} \eta \wedge \rho} & =&  \frac{M_{1}}{M_{2}}~,
\eea
as in (\ref{ratios}). On the other hand, we may compute the ratio on the left hand side explicitly in terms of toric data using (\ref{Xpqhom}) and the general formula (\ref{toric3}). 
This leads to the following expression for $b_2$ in terms of the 
flux integers $M_{1}$ and $M_{2}$:
\bea
b_{2} & =& -\frac{ b_{1} M_{1} p(M_{1}+ 2 M_{2} )}{M_{1}^2 (p-1) p +2 M_{1}M_{2}p(p-1) +M_{2}^2 [p(p-1)-q(q-1)]}~.
\eea
Finally, setting $b_1=2$ so that the holomorphic volume form has charge 2, the geometric formula for the central charge is
\bea\label{cXpq}
 \cZ  |_\mathrm{on-shell} & =& \tfrac{24 N M_{1}M_{2} (M_{1}+2 M_{2}) p (p-q)(p+q-1)[M_{1} p+ M_{2} (p+q-1)][M_{1} p+M_{2} (p+q)]}{ [(M_{1}+M_{2})^2 (p-1)p+M_{2}^2 q (1-q)]^2}~.
\eea
Note that in this case there is no extremization to do: flux quantization uniquely fixes the R-symmetry vector. 

We may now consider the four-dimensional $X^{p,q}$ quiver gauge theories on $T^2$. 
The theory has $SU(N)^{2p+1}$ gauge group and the field content is summarized in
table \ref{table3}, where the $\lambda$ are the gauginos and the remaining fields are bifundamental matter fields.
\begin{table}[h!]
\small{
\begin{center}
\begin{tabular}{|c|c|c|c|c|c|c|}
\hline
Field&Multiplicity&$R_0$-charge&$U(1)_{B_{1}}$&$U(1)_{B_{2}}$&$U(1)_{F_{1}}$&$U(1)_{F_{2}}$\\
\hline\hline
$X_{12}$&$p N^2$             &$0$   &$0$      &$-1$         &$1$&0\\
$X_{23}$&$(p+q-1)N^2$   &$2$   &$1$      & $2$          &$0$&0\\
$X_{34}$&$N^2$                &$0$   &$-p$     &$-p-q$     &$0$&$1$\\
$X_{45}$&$N^2$                &$0$   &$p$      &$p+q-1$  &$0$&$-1$\\
$X_{51}$&$(p-q)N^2$        &$0$   &$-1$     &$0$          &$-1$&0\\
$X_{24}$&$N^2$                &$2$   &$1-p$   &$2-p-q$   &$0$&$1$\\
$X_{31}$&$(q-1)N^2$        &$0$   &$-1$     &$-1$          &$-1$&$0$\\
$X_{35}$&$(p-1)N^2$        &$0$   &$0$      &$-1$          &0&0\\
$X_{41}$&$N^2$                &$0$   &$p-1$   &$p+q-1$  &$-1$&$-1$\\
$X_{52}$&$q N^2$             &$0$   &$-1$     &$-1$         &0&0\\
$\lambda$&$(2p+1)(N^2-1)$&1&0&0&0&0\\
\hline
\end{tabular}
\caption{The field content of the $X^{p,q}$ quiver theories.}
\label{table3}
\end{center}
}
\end{table}
Here $U(1)_{B_I}$, $I=1,2$, correspond to baryonic symmetries associated to the three-cycles 
$\sigma_1$, $\sigma_2$, respectively, while $U(1)_{F_i}$, $i=1,2$, are flavour symmetries corresponding to $U(1)$ isometries 
under which the holomorphic volume form is uncharged. 

We wrap these theories on $T^2$, introducing only baryonic flux in the topological twist
\bea
T_{\mathrm{top}} & =&  \betabeta_{1} T_{B_{1}}+\betabeta_{2} T_{B_{2}}~.
\eea
Geometrically this corresponds to a product $T^2\times Y_5$. The trial R-charge is a linear combination
\bea
T_{\text{trial}} &=&  T_{R_0}+\zeta_{1} T_{B_{1}}+\zeta_{2} T_{B_{2}}+ \epsilon_{1} T_{F_{1}}+\epsilon_{2} T_{F_{2}}~,
\eea
where $\zeta_I$, $\epsilon_i$ are parameters. Substituting this data into the trial central charge 
of \cite{Benini:2012cz} and extremizing over the parameters $\zeta_I$, $\epsilon_i$, we find
\bea\label{cextXpq}
c_{\text{c-ext}} & =& \tfrac{24 N^2\betabeta_{1}\betabeta_{2} (\betabeta_{1}+2 \betabeta_{2})  p (p-q)(p+q-1)([\betabeta_{1} p+ \betabeta_{2} (p+q-1)][\betabeta_{1} p+\betabeta_{2} (p+q)]}{ [(\betabeta_{1}+\betabeta_{2})^2 (p-1)p+\betabeta_{2}^2 q (1-q)]^2}~.
\eea
This agrees with the geometric result (\ref{cXpq})
on making the identification
\bea
M_{a} &=&  N \betabeta_{a}~,
\eea
of geometric and field theory baryonic flux parameters $M_a$, $\betabeta_a$, respectively.
As for the quiver gauge theories for $Y^{p,q}$ and $L^{a,b.c}$ that we discussed in previous subsections,
we can also determine the R-charges of various fields, and a numerical investigation with $q<p\le 500$ shows that
there is always a chiral operator with negative R-charge. Thus, we can again conclude that the associated 
superconformal fixed point with central charge as in \eqref{cextXpq} with $p>q>0$ does not exist. It again remains an open problem
as to the fate of these quiver gauge theories compactified on $T^2$ with baryonic flux only,
a point we return to in section
\ref{sec:discuss}.

Notice, however, that (\ref{cXpq}) 
is positive if $q>p>0$ and $M_1,M_2>0$. In fact we find 
that all the R-charges of fields are also positive in this range, at least for $p< q\le 500$. Given the similarities with the explicit $\Ypq$ and $\Labc$ 
solutions in section \ref{sec:Ypq} and \ref{sec:Labc}, we are thus naturally led to conjecture 
that the corresponding complex cones admit compatible AdS$_3\times T^2\times \mathscr{X}^{p,q}$ supergravity solutions with $q>p>0$.
These complex cones have the same toric data as (\ref{Xpqv}), but with $q>p>0$; they are non-convex, much like the 
$\Ypq$ and $\Labc$ examples. Unlike those examples, since we do not have explicit metrics, here we need to assume that
our conjecture that (\ref{toric3}) and (\ref{toricconstraint}) are valid in order to calculate the central charge and the R-charges.
Assuming these full solutions exist they will necessarily be cohomogeneity two, with no expectation that the equation of motion can be solved by separation 
of variables, so finding these solutions explicitly would involve solving a non-linear PDE in two variables. 
Again, we have no proposal for the dual field theory description. 

\subsection{$Y_5= Z^{p,q}$ }\label{sec:Zpq}

In the previous section we have shown that for $b_{3}(Y_{5})=2$ it is not necessary to extremize the trial central charge in the geometric computation as this is determined uniquely by flux quantization. Instead consider the $Z^{p,q}$ quiver theories of \cite{Oota:2006eg}. These have $b_{3}(Z^{p,q})=3$ and we shall see that the gravity result not only fixes the central charge uniquely by flux quantization, but also fixes one of the flux quantum numbers $M_{I}$. These field theories blow down to the $X^{p,q}$ theories considered in the previous section, and contain the dP$_3$ theory as a special limit ($p=2, ~q=1$).

As before the Calabi-Yau cone over the $Z^{p,q}$ manifold is toric, where $p\geq q>0$ are integers. The inward-pointing normal vectors are
\bea \label{Zpqv}
\v_1&=&(1,1,p) ~, \quad \v_2\ =\ (1,0,p-q+1)~  , \quad \v_3\ =\ (1,0,p-q)~ ,\nonumber\\
\v_4 &=&(1,1,0)~, \quad \v_5 \ =\ (1,2,0)~ , \quad \v_6\ =\ (1,2,1)~.
\eea
Using the toric data \eqref{Zpqv} we may apply our general formulae from section \ref{sec:examplesT2}. To prevent expressions becoming completely unwieldy, below we shall only present explicit results for dP$_3$, that is $p=2$, $q=1$. However, it is straightforward to compute for general $p$ and 
$q$ using a computer algebra package. Thus setting $p=2$, $q=1$, using  (\ref{toricconstraint}) the
constraint equation (\ref{constraint5}) implies
\be
0\ =\ -\frac{2 b_{1}^{2}(b_{2}^2 + b_{2}b_{3}+b_{3}^2 -3 b_{1}(b_{2}+b_{3}))}{b_{2}b_{3}(2 b_{1}-b_{2})(2 b_{1}-b_{3})(b_{1}-b_{2}-b_{3})(b_{2}+b_{3}-3 b_{1})}~,
\ee
which admits the solution
\be
b_{3} \ = \ \frac{1}{2}\left(3 b_{1}-b_{2}+\sqrt{9 b_{1}^2 +6b_{1}b_{2}-3 b_{2}^2}\right)~.
\ee
There are six toric divisors which map to the facets $\{ (\vec{y}, \v_{a})=0\}$ in the moment map cone $\mathcal{C}\, , a=1,..,6$. These give rise to cones over torus-invariant three-dimensional submanifolds $S_{a}$. Since $b_{3}(Z^{p,q})=3$ there are three generating three-cycles which we call $\sigma_{1},\ \sigma_{2},\ \sigma_{3}$. The homology relations in $H_{3}(Z^{p,q}, \Z)$ are (temporarily restoring general $p$ and $q$)
\bea\label{ZpqHom}
[S_{1}] &=& (p-q) \sigma_{1}~, \quad  [S_{2}]\ =\ (p-q)\sigma_{2}~, \quad [S_{3}]\ = \ -p \sigma_{1}+(q-p-1) \sigma_{2}-\sigma_{3}~, \quad \nn\\
 \left[S_{4}\right]& =& (p+q)\sigma_{1}+2 \sigma_{2} +2 \sigma_{3} ~, \quad \left[ S_{5} \right] \ =  \ -p \sigma_{1}- \sigma_{2} +(q-p-1) \sigma_{3}~ ,\nn\\ \quad [S_{6}] & = & (p-q)\sigma_{3}~.  
\eea
Analogously to the previous section, flux quantization imposes
\bea
\int_{\sigma_{1}} \eta \wedge \rho & = & \frac{(2 \pi \ell_{s})^4 g_{s}}{A L^4} M_{1} ~,\quad
\int_{\sigma_{2}} \eta \wedge \rho \ =\  \frac{(2 \pi \ell_{s})^4 g_{s}}{A L^4}M_{2} ~,\nn\\
\int_{\sigma_{3}} \eta \wedge \rho & = & \frac{(2 \pi \ell_{s})^4 g_{s}}{A L^4}M_{3}~,
\eea
which implies the \emph{two} conditions
\be
\frac{\int_{\sigma_{1}} \eta \wedge \rho }{\int_{\sigma_{2}} \eta \wedge \rho} \ =\  \frac{M_{1}}{M_{2}}~,\qquad \frac{\int_{\sigma_{1}} \eta \wedge \rho }{\int_{\sigma_{3}} \eta \wedge \rho} \ = \ \frac{M_{1}}{M_{3}}~.
\ee
We may compute the ratio on the left hand side of each expression explicitly in terms of toric data by using \eqref{ZpqHom} and formula \eqref{toric3}. The first condition fixes $b_{2}$ in terms of the flux integers $M_{1}, M_{2}$ as in the $X^{p,q}$ case 
\be
b_{2} \ = \ \frac{3b_{1}M_{1}(M_{1}+M_{2})}{M_{1}^2+M_{1}M_{2} +M_{2}^2}~,
\ee
whilst the second places a restriction on the possible flux integers
\be\label{dP3Mrelation}
M_{3} \ = \ -\frac{M_{1}(2 M_{1}+  M_{2})}{M_{1}-M_{2}}~.
\ee
Setting $b_{1}=2$ the geometric central charge for $p=2$, $q=1$ is
\bea
 \cZ  |_\mathrm{on-shell}  & =& \frac{72 N M_{1}M_{2}(M_{1}+M_{2})(2M_{1}+M_{2})(M_{1}+2 M_{2})}{(M_{1}^2+M_{1}M_{2}+M_{2}^2)^2}~.
\eea
We have seen that not only is there no extremization to do, but also that only certain twists are possible, \eqref{dP3Mrelation}. Moreover, flux quantization implies that the $M_{I}$ are integer and therefore only $M_{1}$ and $M_{2}$ such that \eqref{dP3Mrelation} is integer are permissible. Note that there are an infinite number of choices of $M_1$ and $M_2$ which give an integer result for (\ref{dP3Mrelation}).

Let us now consider the four-dimensional $Z^{p,q}$ quiver theory reduced on $T^2$ with baryonic flux. The theory is a $SU(N)^{2(p+1)}$ gauge theory with gauginos, $\lambda$ and bifundamental fields transforming under the global symmetries as summarized in table~\ref{table4}.
\begin{table}[h!]
\small{
\begin{center}
\begin{tabular}{|c|c|c|c|c|c|c|c|c|}
\hline
Field&Multiplicity&$R_{0}$-charge&$U(1)_{B_{1}}$&$U(1)_{B_{2}}$&$U(1)_{B_{3}}$&$U(1)_{F_{1}}$&$U(1)_{F_{2}}$\\
\hline\hline
$X_{61}$&    $(p+q-2)N^2$      & $2$    & $ q-p$     &$ 0$           &$0$            & $0$      &$0$\\
$X_{12}$&    $N^2$                 &$0$     & $0$         &$ q-p$         &$0$            & $0$     &$0$\\
$X_{23}$&    $N^2$                 &$0$     & $ p$        &$ p-q+1$     &$1$            & $0$     &$0$\\
$X_{34}$&    $(p-q)N^2$          &$0$     & $ -p-q$   &$ -2$           &$-2$           & $-1$    &$0$\\
$X_{45}$&    $N^2$                 &$0$     & $ p$        &$ 1$            &$p-q+1$     & $1$     &$-1$\\
$X_{56}$&    $N^2$                 &$0$     & $ 0$        &$ 0$            &$q-p$         & $0$     &$1$\\
$X_{13}$&    $(p-1)N^2$          &$0$     & $ p$        &$ 1$            &$1$           & $0$     &$0$\\
$X_{14}$&    $(q-1)N^2$          &$0$     & $ -q$       &$ -1$           &$-1$          & $-1$    &$0$\\
$X_{24}$&    $N^2$                 &$0$     & $ -q$       &$ p-q-1$      &$-1$          & $-1$    &$0$\\
$X_{35}$&    $N^2$                 &$0$     & $ -q$       &$ -1$           &$p-q-1$     & $0$      &$-1$\\
$X_{36}$&    $(q-1)N^2$          &$0$     & $ -q$       &$ -1$           &$-1$          & $0$     &$0$\\
$X_{46}$&    $(p-1)N^2$          &$0$     & $ p$        &$ 1$            &$1$            & $1$     &$0$\\
$X_{51}$&    $N^2$                 &$2$     & $ q-p$     &$ 0$             &$q-p$        & $0$     &$1$\\
$X_{62}$&    $N^2$                 &$2$     & $q-p$      &$ q-p$          &$0$           & $0$     &$0$\\
$\lambda$&$2(p+1)(N^2-1)$&$1$&$0$&$0$& $0$&$0$&$0$\\
\hline
\end{tabular}
\caption{The field content of the $Z^{p,q}$ quiver theories.}
\label{table4}
\end{center}
}
\end{table}
As before $U(1)_{B_{I}}, I=1,2,3$ are the baryonic symmetries associated to the three three-cycles $\sigma_{I}$ respectively, while $U(1)_{F_{i}}, i=1,2$ are the flavour symmetries. 

We wrap these theories on $T^2$ by introducing only baryonic flux in the topological twist
\be
T_{\mathrm{top}}\ =\ \betabeta_{1} T_{B_{1}}+\betabeta_{2} T_{B_{2}}+\betabeta_{3} T_{B_{3}}~.
\ee
The trial R-charge is the linear combination 
\be
T_{\mathrm{trial}}\ =\  T_{R_{0}}+ \zeta_{1} T_{B_{1}}+\zeta_{2} T_{B_{2}}+\zeta_{3} T_{B_{3}}+\epsilon_{1} T_{F_{1}}+\epsilon_{2} T_{F_{2}}~,
\ee
with $\zeta_{I}$ and $\epsilon_{i}$ parameters over which we extremize. Inserting this data into the trial central charge and extremizing we find that a non-zero solution exists only if
\be
\betabeta_{3} \ = \ -\frac{\betabeta_{1}(2\betabeta_{1}+ \betabeta_{2})}{\betabeta_{1}-\betabeta_{2}}~,
\ee
{\it c.f.} \eqref{dP3Mrelation}. The central charge is then
\be
c_{\mathrm{c-ext}} \ = \ \frac{72 N^2 \betabeta_{1}\betabeta_{2} (\betabeta_{1}+\betabeta_{2})(2 \betabeta_{1}+\betabeta_{2})(\betabeta_{1}+2 \betabeta_{2})}{(\betabeta_{1}^{2}+\betabeta_{1}\betabeta_{2}+\betabeta_{2}^2)^2}~.
\ee
To compare with the gravity result one should set 
\be
M_{I} \ =\  N \betabeta_{I}~,
\ee
which gives perfect agreement. 

Field theoretically one can understand the necessity for the relation between the twist parameters by using the fact that the cubic 't Hooft anomalies for mixed baryonic symmetries vanish for these theories, $k_{B_{I_1}B_{I_2}B_{I_3}}=0$. This implies that the mixing parameters of the baryonic symmetries in the trial R-symmetry, the $\zeta_{I}$'s, appear linearly and not quadratically in the trial central charge. Extremizing over these parameters first necessarily implies $b_{3}(Y_5)$ conditions, where $b_3(Y_5)$ is the third Betti number of the associated Sasaki-Einstein five-manifold. In the case when the number of baryonic symmetries exceeds the number of flavour symmetries $n_{F}$, consistency implies that there are $b_{3}(Y_5)-n_{F}$ relations between the baryonic twist parameters $\betabeta_{I}$. This is the field theoretic analogue of the discussion of the latter part of section \ref{sec:geometry}. Note that this is a peculiarity of the type of topological twist being performed here, and that a more general topological twist including flavour and R-symmetry (that is not on a $T^2$) will not in general have linear $\zeta_{I}$'s in the trial central charge.

Some numerical investigation shows that the putative superconformal filed theory cannot exist, since with
$p>q$ we always seem to find chiral operators with negative R-charge. Thus, once again it remains an open question to
determine the fate of these quiver gauge theories compactified on a $T^2$ with baryonic flux only.
We discuss this below in section
\ref{sec:discuss}.

Finally, it is natural to conjecture that the complex cones $\mathscr{Z}^{p,q}$ defined by (\ref{Zpqv}) with $q>p>0$ admit compatible
supergravity AdS$_3\times T^2\times \mathscr{Z}^{p,q}$ solutions. 
These complex cones have the same toric data as (\ref{Zpqv}), but with $q>p>0$, and are consequently non-convex.
Again, one can check that the central and R-charges are positive in this case, assuming the validity of our conjectured formulae (\ref{toric3}) and (\ref{toricconstraint}) 
for the case of non-convex toric cones.


\section{Discussion}\label{sec:discuss}

Inspired by the geometric dual of $a$-maximization put forward in \cite{Martelli:2005tp}, and elaborated upon in \cite{Martelli:2006yb,Gauntlett:2006vf}, in this paper we have formulated a geometric problem that allows one to determine various properties of a class of odd-dimensional 
``GK geometries" $Y_{2n+1}$ that arise in certain AdS supergravity solutions \cite{Gauntlett:2007ts}. In particular, assuming a solution exists, the  R-symmetry Killing vector on $Y_{2n+1}$
may be determined by extremizing a function that depends only on certain global, topological data, without knowing the explicit 
form of the solution.
In seven dimensions these backgrounds characterize AdS$_3\times Y_7$ solutions of type IIB supergravity \cite{Kim:2005ez}, that are holographically dual to two-dimensional (0,2) SCFTs with a $U(1)_R$-symmetry, and therefore our geometric problem may be interpreted as a dual to the $c$-extremization 
principle \cite{Benini:2012cz} in these theories. In nine dimensions instead the backgrounds characterize AdS$_2\times Y_9$ solutions of eleven-dimensional supergravity that are dual to one-dimensional
SCFTs with two supercharges \cite{Kim:2006qu} and we have shown that our new variational principle
allows one to obtain the two-dimensional Newton constant which is naturally associated with the partition function
of the dual superconformal quantum mechanics. For a sub-class of AdS$_2\times Y_9$ solutions we showed that
the variational principle governs the entropy of a certain class of AdS$_4$ black hole solutions, as well as giving
the twisted topological index of certain $\mathcal{N}=2$ SCFTs compactified on a Riemann surface with a universal twist.

The class of geometries studied in this paper is of independent mathematical interest. The work here, extending that
of \cite{Gauntlett:2007ts}, can be viewed as initial steps in developing a programme analogous to that for Sasakian geometry \cite{Boyer:2008era}. Similarly to the latter, in our case the geometry is 
foliated  by a canonical R-symmetry Killing vector with constant
norm, with the leaf space being locally conformally K\"ahler 
(\emph{c.f.} (\ref{Reeb}) -- (\ref{fixB})), while the associated metric cone in one dimension higher is here  complex, but \emph{not} K\"ahler.  This crucial difference with respect to Sasakian geometry implies that the supersymmetric  geometry, as defined  in this paper, has some distinctive features which will be interesting to explore further. 

In this paper we  specialized in sections \ref{sec:examplesT2} and \ref{sec:examples} to  a sub-class of AdS$_3\times Y_7$ type IIB solutions 
where $Y_7=T^2\times Y_5$. In our examples the five-dimensional geometry of $Y_5$ was toric, 
in the sense of possessing a $U(1)^3$ isometry 
that lifts to a corresponding $(\C^*)^3$ holomorphic action on the complex cone $C(Y_5)\cong \R_{>0}\times Y_5$. However, we have also explained that these are not toric in the sense of symplectic geometry, nor in the usual sense of algebraic  geometry. For this reason this class of  geometries, which we called non-convex toric geometries,
cannot be studied with the standard tools of toric geometry (convex polytopes, {\it etc}), at least not without appropriate modification.
We believe that developing a mathematical framework for these non-convex toric geometries would be worthwhile.
In particular, it would be interesting to prove our conjecture that (\ref{toric3})--(\ref{toricconstraint}) hold not only\
in the toric case, but also in the non-convex toric case. Significant evidence for the validity of this conjecture is that
they give the correct gravitational central charge as well as the R-symmetry charges of certain baryonic operators
for the specific examples of $Y_5=\Ypq$ and $\Labc$, for which explicit solutions are known.
In sections \ref{sec:Xpq}, \ref{sec:Zpq} we have  conjectured the existence of  new classes of AdS$_3$ $\times T^2 \times Y_5$  solutions, 
where $Y_5=\mathscr{X}^{p,q}$, $Y_5=\mathscr{Z}^{p,q}$, respectively, are non-convex toric. 
It is clearly important to understand the necessary and sufficient conditions for the existence of metrics satisfying  (\ref{boxR}), analogous
 to the general existence theorem for toric Sasaki-Einstein metrics in \cite{Futaki:2006cc}.

It is still an  interesting open problem to identify the dual $(0,2)$ SCFTs 
for \emph{any} of the AdS$_3\times T^2\times Y_5$ solutions discussed in section \ref{sec:examples} which, in addition to the above examples, also included $Y_5=\Ypq$ \cite{Donos:2008ug} and a new class of explicit solutions $\Labc$ that generalise the local construction of \cite{Donos:2008hd}.
Conversely, we have argued that taking any four-dimensional quiver gauge theory, dual to an AdS$_5\times SE_5$ Sasaki-Einstein solution, 
and compactifying on $T^2$, does not flow to a corresponding AdS$_3\times T^2\times Y_5$ solution where the complex 
structures of $C(Y_5)$ and $C(SE_5)$ are the same, specifically because this latter solution doesn't exist. 

Furthermore, we have shown that there are fundamental problems in
carrying out the $c$-extremization procedure for various quiver gauge theories associated with Sasaki-Einstein manifolds,
when reduced on $T^2$ with a baryonic twist. For example, for the quiver gauge theories associated with the $Y^{p,q}$ Sasaki-Einstein geometries (with $p>q>0$), demanding that the resulting central charge \eqref{cextYpq} is positive leads to negative R-charges for certain chiral fields -- see \eqref{YpqR}. Following the discussion of \cite{Kutasov:2003iy} in the context of 
$a$-maximization, it is possible that these gauge theories do indeed flow to a $(0,2)$ SCFT in the IR, but that certain operators 
are becoming free along the RG flow and, in addition, that the true R-symmetry also involves an ``accidental" global symmetry that only appears at the IR fixed point. In some cases when this occurs for $d=4$, $\mathcal{N}=1$ SCFTs  one can successfully implement a 
refined version of $a$-maximization where one eliminates the decoupled gauge-invariant operators \cite{Kutasov:2003iy} 
(see also {\it e.g.} \cite{Corrado:2004bz,Maruyoshi:2016tqk,Benvenuti:2017lle}). Moreover, an analogous approach has been used
in the context of $c$-extremization for certain $d=2$, $\mathcal{N}=(0,2)$ Landau-Ginzburg theories \cite{Bertolini:2014ela,Gadde:2016khg}. In the $Y^{p,q}$ examples considered here, from \eqref{YpqR} we see that an infinite tower of gauge-invariant chiral operators, built from bifundamental chiral fields $Y$ and $Z$, need to be decoupled. 
We do not know of any 
examples where such fields have been decoupled, either in the context of $a$-maximization or $c$-extremization, making this a particularly interesting avenue to investigate further.
We emphasize that successfully carrying out such a refined $c$-extremization is  unlikely to modify our conclusion
that the resulting $\mathcal{N}=(0,2)$ SCFT is not dual to the relevant AdS$_3\times T^2\times Y_5$ solutions, as we have mentioned several times. Specifically, in
the $Y^{p,q}$ quiver gauge theory we have $p>q>0$ while we have $q>p>0$ in the  AdS$_3\times T^2\times \Ypq$ solutions.

We conclude by mentioning some other directions for future work. Based on the examples that we have analysed, it is natural to conjecture that 
the holographic central charge $\cZ$, as a function of the geometric data, agrees with the field theory central charge 
$c_R$. Of course, in some sense this is just a restatement of what the AdS/CFT correspondence conjectures, but we have seen that this relation holds 
 for classes of four-dimensional field theories compactified on $T^2$, even 
when there is no corresponding supergravity solution. There hence seems to be a stronger mathematical identity at work,  possibly holding off-shell similarly to 
\cite{Butti:2005vn}, that will imply the expected AdS/CFT relation when solutions to the supergravity equations do exist.
It would be interesting to try to prove this claim, perhaps using the observations made in \cite{Amariti:2017iuz}. 
We also note that since the trial $c$-function of \cite{Benini:2012cz} is quadratic, the superconformal R-symmetry 
and central charge $c_R$ of a superconformal $(0,2)$ theory should be rational. This fact is not immediate from our general geometric extremal 
problem summarized
in section \ref{sec:c}, although 
we note that it is true in all of the examples we have analysed. Understanding the precise relation between 
our off-shell geometric central charge $\mathscr{Z}$ and the trial $c$-function, perhaps along the lines of \cite{Butti:2005ps} relating 
$Z$-minimization and $a$-maximization, might also help to clarify this issue.

In a different direction, an obvious continuation of our work is to analyse cases where the seven-dimensional manifolds $Y_7$ are not of the form  $T^2 \times Y_5$. 
It would be interesting to further develop our understanding of the geometry, including the toric case where it is known 
that the toric examples in the class \cite{Gauntlett:2006af} also have cones with non-convex toric geometries \cite{gauntkimwald}. 
Finally, and similarly to how the results of \cite{Martelli:2005tp}
predicted  ${\cal F}$-extremization in $d=3$, ${\cal N}=2$ SCFTs, 
our results also strongly suggest that there exists a general extremization principle for ${\cal N}=2$ superconformal quantum mechanics
with a $U(1)$ R-symmetry, extending the proposal of \cite{Benini:2015eyy}. 

\subsection*{Acknowledgments}
We would like to thank Seok Kim, Daniel Waldram and Alberto Zaffaroni for helpful discussions.
CC was supported by an STFC studentship under the STFC rolling grant ST/N504361/1 and also
acknowledges the support of the Netherlands Organization for Scientifc Research (NWO) under the VICI grant 680-47-602.
JPG is supported by the European Research Council under the European Union's Seventh Framework Programme (FP7/2007-2013), ERC Grant agreement ADG 339140. JPG is also supported by STFC grant ST/P000762/1, EPSRC grant EP/K034456/1, as a KIAS Scholar and as a Visiting Fellow at the Perimeter Institute.  
DM is supported by the ERC Starting Grant 304806 ``The gauge/gravity duality and geometry in string theory''.


\appendix

\section{Black hole entropy and the two-dimensional Newton constant}\label{norm}
The $D=4$ magnetically charged black hole of interest is given by \cite{Caldarelli:1998hg}
\begin{align}
\diff s^2_4& \ = \ -\left(\rho-\frac{1}{2\rho}\right)^2\diff t^2+\left(\rho-\frac{1}{2\rho}\right)^{-2}\diff\rho^2+\rho^2 \diff s^2(H_2)~,\nn\\
F& \ = \ \vol(H_2)~.
\end{align}
It is a solution of minimal $D=4$ gauged supergravity with bosonic action
\begin{align}
I \ = \ \frac{1}{16\pi G_4}\int \diff ^4 x\sqrt{-g}\left(R+6-\frac{1}{4}F^2\right)~,
\end{align}
and has Bekenstein-Hawking entropy given by
\begin{align}\label{sbh}
S_{BH}
\ = \ \frac{\vol(H_2)}{8G_4}~.
\end{align}
We can uplift this on an arbitrary seven-dimensional Sasaki-Einstein manifold, $SE_7$, 
to obtain a solution of $D=11$ supergravity using (2.1) of \cite{Gauntlett:2007ma}. We can write the
metric on $SE_7$ as 
$\diff s^2(SE_7)=  (\diff \psi+\sigma)^2+\diff s^2(KE_6)$ with $\diff \sigma=2J_{KE}$ and $\rho(KE_6)= 8J_{KE_6}$. 
Choosing a convenient 
overall length normalization, the $D=11$ metric is given by
\begin{align}\label{gen11}
\diff s^2_{11}\ = \ \frac{4L^2}{2^{2/3}}\left[\diff s^2_4+4\left((\diff\psi+\sigma+\tfrac{1}{4}A)^2+\diff s^2(KE_6)\right)\right]~,
\end{align}
where $F=\diff A$.
With this length normalization we see that as a $D=11$ metric the black hole horizon metric is given by
\begin{align}\label{gen112}
\diff s^2_{11}& \ = \ \frac{L^2}{2^{2/3}}\left[\diff s^2(\mathrm{AdS}_2)+2\diff s^2(H^2)+(\diff z+P)^2+16 \diff s^2(KE_6)\right]~,
\end{align}
where we have rescaled $z=4\psi$ and $P=4\sigma+A$. This is precisely the metric 
of the AdS$_2$ solution constructed directly as in (6.15) of \cite{Gauntlett:2006ns}, and with length normalization
as in this paper.
Now, using the $D=11$ metric \eqref{gen11}, we can reduce to $D=4$ and deduce that
\begin{align}
\frac{1}{G_4} \ = \ \frac{1}{G_{11}}2^{13}L^9\vol(SE_7)~.
\end{align}
On the other hand, using \eqref{gen112} we can reduce to two dimensions to obtain
\begin{align}
\frac{1}{G_2} \ = \ \frac{1}{G_{11}}2^{12}L^9\vol(SE_7)\vol(H_2)~.
\end{align}
Combined with \eqref{sbh} we conclude that
\begin{align}
S_{BH} \ = \ \frac{1}{4G_2}~.
\end{align}

\section{Relating AdS$_2\times T^2\times Y_7$ and AdS$_3\times Y_7$ solutions}\label{dimredtdual}
Consider an AdS$_2\times Y_9$ solution in $D=11$ with $Y_9=T^2\times Y_7$. This is related by dimensional reduction
on one leg of the $T^2$ followed by  T-duality on the other leg to an AdS$_3\times Y_7$ solution of type IIB, as shown in appendix C of \cite{Gauntlett:2006qw}. Here we make a precise connection between the two-dimensional Newton 
constant, $G_2$, and the central charge of the $d=2$ SCFT, $\csugra$, which were defined in section \ref{sec:c}.

We first recall some well-known results. We assume that the torus metric in $D=11$ is given by $(\diff x^9)^2+(\diff x^{10})^2$,
with periodic coordinates 
$(x^9,x^{10})=(x^9+2\pi R_2,x^{10}+2\pi R_1)$. We first reduce along $x^{10}$, where we have $R_1\equiv \ell_s g_s^{IIA}$.
By integrating the $D=11$ supergravity action over $x^{10}$ and relating it to the type IIA action we deduce $\frac{R_1}{\ell_p^9}=\frac{1}{\ell^8_s(g_s^{IIA})^2}$.
We next carry out a T-duality over the $x^9$ direction and we note that $R_2$ is the type IIA radius of this circle. We then have the
T-duality formulae for the type IIB radius, $R_{IIB}=\frac{\ell_s^2}{R_2}$, and the type IIB string coupling, $g_s=\frac{g_s^{IIA}\ell_s}{R_2}$. From this we conclude the following relations between the type IIB quantities and the $D=11$ quantities:
\begin{align}\label{gendualform}
l_s^2\ = \ \frac{l_p^3}{R_1}~,\qquad g_s \ = \ \frac{R_1}{R_2}~,\qquad R_{IIB}\ = \ \frac{\ell_p^3}{R_1R_2}~.
\end{align}

The $D=11$ solution of interest has the form
\begin{align}
\diff s^2_{11}&\ = \ L_{11}^2\ex^{-2B_{11}/3}[\diff s^2(\mathrm{AdS}_2)+(\diff z+P)^2+\ex^{B_{11}}\diff s^2_6]+\ex^{B_{11}/3}L_{11}^2\diff s^2( T^2)~,\nn\\
*_{11} G& \ = \  L_{11}^{6} \left[ (\dd z +P) \wedge \rho_6 \wedge \left( \frac{J_6^2}{2} + J_6\wedge \vol_{T^2}\right) + \frac{1}{2} * \dd R_6\right]~,
\end{align}
where we have added some subscripts for clarity. We make the identification 
\begin{align}\label{elelsq}
L_{11}^2\vol(T^2)\ = \ (2\pi)^2R_1R_2~.
\end{align}
We now use the dimensional reduction and T-duality formula given, for example, in appendix C of \cite{Gauntlett:2006qw} to obtain a type IIB solution whose metric is given by
\begin{align}
\diff s^2_{10}& \ = \ L_{11}^2\ex^{-B_{11}/2}[\diff s^2(\mathrm{AdS}_2)+(\diff z+P)^2+\ex^{B_{11}}\diff s^2_6]+\ex^{-B_{11}/2}L_{11}^2(\diff \phi+a_1)^2~,
\end{align}
with $\diff a_1=\vol(\mathrm{AdS}_2)$. In the type IIB solution we now have 
\begin{align}\label{delxpiib}
\Delta  (L_{11} \phi) \ = \ 2\pi R_{IIB}~.
\end{align}
Recalling that we also have $\ex^{B_{11}}=R_6/2=4\ex^{B_{10}}$, we can rewrite the metric in the form
\begin{align}\label{thecompadsth}
\diff s^2_{10}&\ = \ (2L_{11}^2)\ex^{-B_{10}/2}\left[\frac{1}{4}\diff s^2(\mathrm{AdS}_2)+\frac{1}{4}(\diff \phi+a_1)^2+\frac{1}{4}(\diff z+P)^2+\ex^{B_{10}}\diff s^2_6\right] \nn\\
&\ =(2L_{11}^2)\ex^{-B_{10}/2}\left[\diff s^2(\mathrm{AdS}_3)+\frac{1}{4}(\diff z+P)^2+\ex^{B_{10}}\diff s^2_6\right]~,
\end{align}
which is now precisely in the form of the type IIB AdS$_3$ solutions, in the notation of this paper, provided that
we take 
\begin{align}\label{lsqrel}
L_{10} \ = \ 2^{1/2} L_{11}~.
\end{align}

With these results in hand, we can now relate the two-dimensional Newton constant, $G_2$, to the
one in three dimensions, $G_3$, and hence the central charge, $\csugra$, of the $d=2$ SCFT. Starting with
\eqref{G2gen} and using \eqref{cSUGRAgen} we deduce that
\be
\frac{1}{G_2} \ = \ 8\frac{L_{11}^9G_{10}}{L_{10}^7G_{11}}\vol(T^2)\frac{1}{G_3}~.
\ee
Next, using \eqref{ncten}, \eqref{ncel}, \eqref{gendualform}, \eqref{elelsq} and \eqref{delxpiib} we can conclude that
\begin{align}\label{result}
\frac{1}{G_2}&\ = \ \frac{2\pi}{3}\frac{\Delta\phi}{2\pi}\csugra~.
\end{align}

We now return to the AdS$_2\times T^2\times Y_7$ solution in $D=11$ and consider the quantization of $*_{11}G$. 
There are two types of seven-cycles to analyse: the product of a five-cycle $\Sigma_A^{(5)}$ in $Y_7$ and the $T^2$, and $Y_7$ itself.
For the former, we have
\begin{align}
N^{(11)}_A \ \equiv \ \frac{1}{(2 \pi \ell_{p})^6}\int_{\Sigma_{A}^{(5)}\times T^2} *_{11} G
&\ = \ \frac{2 L_{11}^6 \vol({T}^2)}{(2 \pi \ell_p)^6} \int_{\Sigma^{(5)}_{A}}\frac{1}{2} (\dd z+P) \wedge \rho_6 \wedge J_6 \nn\\
& \ = \ 4 \frac{L_{11}^4}{L_{10}^4} \frac{L_{11}^2\vol({T}^2)}{(2\pi)^2} \frac{\ell_s^4 g_s}{\ell_p^6} N_{A}^{IIB}\nn\\
& \ = \ N_{A}^{(IIB)}~,
\end{align}
where we have used the definition of the type IIB flux quantization condition \eqref{quantize} 
in the second line and then \eqref{gendualform}, \eqref{elelsq} and \eqref{lsqrel} to get to the third. In other words, 
the four-form flux quantization in $D=11$ for these seven-cycles is equivalent to the five-form flux quantization condition in type IIB for the five-cycles in $Y_7$. 

We now turn to the quantization condition for the second type of seven-cycle. We calculate as follows:
\begin{align}\label{intcond}
N^{(11)}\ \equiv\ \frac{1}{(2\pi\ell_p)^6}\int_{Y_7} *_{11} G&\ = \ \frac{L_{11}^6}{(2 \pi \ell_p)^6} \int_{Y_7}(\dd z+P) \wedge \rho_6 \wedge \frac{J_6^2}{2} \nn\\
&\ = \ \frac{2 L_{11}^6}{(2 \pi \ell_p)^6}\frac{(2 \pi)^6 \ell_s^8 g_s^2 }{3 L_{10}^8}c_{\text{sugra}}\nn\\
&\ =\ \frac{c_{\text{sugra}}}{24} \left(\frac{\Delta\phi}{2\pi}\right)^2 \in \mathbb{Z}~,
\end{align}
where we used (\ref{Ssusy}),\eqref{cS} and \eqref{cS2} to get the second line. 

One way to read these relations is as follows. Start with a {\it bona fide} AdS$_2\times T^2\times Y_7$ solution,{ \it  i.e.} with properly quantized four-form flux. In particular it has specific values of $\frac{1}{G_2}$ and $N^{(11)}$. Then, after eliminating $\csugra$ from \eqref{result} and \eqref{intcond}, we see that after dimensional reduction and T-duality
we obtain a {\it bona fide} AdS$_3\times Y_7$ solution, with 
AdS$_3$ metric as in \eqref{thecompadsth}, with $\phi$ having specific period given by
\begin{align}
\frac{\Delta\phi}{2\pi}\ = \ {16\pi}{G_2}N^{(11)}~.
\end{align}

Alternatively, start with a {\it bona fide}  AdS$_3\times Y_7$ solution with, in particular, a specific value of $\csugra$. Then, writing the AdS$_3$ metric as in \eqref{thecompadsth} and demanding that $\phi$ has period $\Delta\phi$, then after 
T-duality and uplifting we obtain a {\it bona fide}  AdS$_2\times T^2\times Y_7$ solution only if
\eqref{intcond} is satisfied. 


\providecommand{\href}[2]{#2}\begingroup\raggedright\endgroup

\end{document}